\documentclass[english,aps,manuscript,showpacs]{revtex4}
\usepackage[T1]{fontenc}
\usepackage[latin9]{inputenc}
\usepackage{color}
\usepackage{amsmath}
\usepackage{amssymb}
\usepackage{graphicx}
\usepackage{esint}

\makeatletter
\@ifundefined{textcolor}{}
{%
 \definecolor{BLACK}{gray}{0}
 \definecolor{WHITE}{gray}{1}
 \definecolor{RED}{rgb}{1,0,0}
 \definecolor{GREEN}{rgb}{0,1,0}
 \definecolor{BLUE}{rgb}{0,0,1}
 \definecolor{CYAN}{cmyk}{1,0,0,0}
 \definecolor{MAGENTA}{cmyk}{0,1,0,0}
 \definecolor{YELLOW}{cmyk}{0,0,1,0}
 }

\makeatother

\usepackage{babel}
\begin{document}

\title{Protecting a quantum state from environmental noise by an incompatible
finite-time measurement}

\author{Carlos Alexandre Brasil}

\email{carlosbrasil@ifsc.usp.br}

\author{L. A. de Castro}

\email{leonardo.castro@usp.br}

\author{R. d. J. Napolitano}

\affiliation{Instituto de Física de São Carlos, Universidade de São Paulo, P.O.
Box 369, 13560-970, São Carlos, SP, Brazil}
\begin{abstract}
We show that measurements of finite duration performed on an open
two-state system can protect the initial state from a phase-noisy
environment, provided the measured observable does not commute with
the perturbing interaction. When the measured observable commutes
with the environmental interaction, the finite-duration measurement
accelerates the rate of decoherence induced by the phase noise. For
the description of the measurement of an observable that is incompatible
with the interaction between system and environment, we have found
an approximate analytical expression, valid at zero temperature and
weak coupling with the measuring device. We have tested the validity
of the analytical predictions against an exact numerical approach,
based on the superoperator-splitting method, that confirms the protection
of the initial state of the system. When the coupling between the
system and the measuring apparatus increases beyond the range of validity
of the analytical approximation, the initial state is still protected
by the finite-time measurement, according with the exact numerical
calculations.
\end{abstract}

\maketitle

\section{Introduction}

The problem of measurement is fundamental to quantum theory \cite{key-1,key-2,key-3,key-4,key-31}.
According to the standard interpretation \cite{key-1,key-3,key-4},
established after the pioneer contributions by Born \cite{key-3,key-44,key-45},
Dirac \cite{key-46}, and von Neumann \cite{key-3}, the postulated
collapse of the wave function implies an instantaneous measurement
process. However, the de facto meaning of a quantum measurement is
the result of a physical interaction between the system being measured
and the measuring apparatus. Such an interaction can be described
by a suitable formalism \cite{key-5,key-6}, which, under the hypothesis
that the measurement process is irreversible, leads to the Lindblad
equation \cite{key-7,key-33,key-34}. This equation describes the
evolution of a measurement during a finite time interval by the stochastic
term - Lindbladian. At last, the information of the probabilities
associated to every possible result is acquired by the conventional
way with the density operator diagonal elements - populations \cite{key-8,key-9}.

Based on superoperator algebra and Nakajima-Zwanzig projectors \cite{key-11,key-12},
we have been able to describe the dynamics of an arbitrary measurement
that occurs during a finite time interval, while the system being
measured interacts with the rest of the universe and, due to the consequent
environmentally-induced noise, undergoes decoherence \cite{key-10}.
The assumption that the interaction of the measuring apparatus with
the system is Markovian justifies a Lindbladian approach. However,
to treat the noise introduced by the fact that, during the finite-duration
measurement, the system is perturbed by the environment, a Markovian
approximation is too restrictive, since non-Markovian noise effects
can become non-negligible for time scales in which coarse graining
is not a good approximation \cite{key-16,key-17,key-18,key-19,key-36,key-39}.
Instead, to develop a formalism able to include non-Markovian effects,
we have used a Redfield approach to the dynamical description of the
interaction between the system and its environment, excluding the
apparatus. After tracing out the degrees of freedom introduced to
describe the non-Markovian noise, the resulting master equation in
the Born approximation (not a Born-Markov approximation), referred
only to the system, can be used to investigate the effects of the
environmental noise on the measurement dynamics. We do not use additional
Lindblad channels to describe the environmental noise because such
a formulation would preclude any attempt to treat non-Markovian effects.
Since we are interested in describing the effects of the noise during
the finite-time measurement, turning off the environmental perturbation
during the action of the measuring apparatus is not a valid approximation.
Therefore, our hybrid approach is the most economical dynamical description
of a measurement that is simultaneous to non-Markovian environmental
perturbations.

The Lindblad equation can be used to describe Markovian and non-Markovian
environmental effects, with applications ranging from decoherence
and dissipation analyses \cite{key-9,key-33} to measurement processes
\cite{key-5}, particularly when we consider that the measurements
have finite durations \cite{key-8,key-9}. It follows as a mathematical
consequence of a semi-group dynamics to describe irreversibility in
quantum mechanics \cite{key-7}. Therefore, it can only be used to
describe Markovian processes, unless the number of dynamical variables
describing the system be enlarged \cite{key-36}, causing computational
overhead that we must avoid. Incidentally, we point out that the extension
of the Lindblad equation to non-Markovian systems \cite{key-36} can
be applied to non-semigroup dynamics as well, with time-dependent
Hamiltonian and Lindbladian operators.

In the present paper, we analyze finite-time measurements that commute
or do not commute with the interaction Hamiltonian. We study a two-state
system interacting with a bath of harmonic oscillators, that emulates
a phase-noisy environment. The approximate analytical solution agrees
with the numerical results of the superoperator-splitting method \cite{key-41}
for weak coupling between the system and its environment. We find
that the finite-time measurement can protect the measured state, if
the observable being measured does not commute with the Hamiltonian
describing the noisy environmental interaction with the system but,
when the measurement commutes with the interaction, the environment
only increases the coherence decay. For strong coupling between the
system and the environment, our analytical solution fails, as shown
by a thorough comparison of its predictions against the corresponding
exact numerical results. In the case of strong coupling, the effect
predicted by the numerical calculations is a more intense error as
the coupling with environment increases.

Subsequent researchers may use the method presented here to study
the effects of varying the temperature, the density of states, or
even the system considered, increasing the number of quantum states.
There are some interesting possible directions to follow to extend
the present investigation, such as the study of non-Markovian systems
\cite{key-16,key-17,key-18,key-19,key-36,key-39}, time-dependent
Nakajima-Zwanzig projectors \cite{key-20,key-30}, higher-order approximations
for the interaction between the system and its environment \cite{key-37,key-38},
and effects of other kinds of environmental noise. One problem which
the present theory can not address is the possibility that the measurement
result be different from one of the eigenvalues of the meter \cite{key-40}.

The paper is structured as follows: In Sec. II we briefly present
the proposed formalism and state the problem to be analyzed. In Sec.
III, we present the approximate analytical solution to the master
equation, followed by a derivation of its numerical counterpart in
Sec. IV. In Sec. V, we discuss the results obtained with the expressions
in Sec. III and Sec. IV and its physical implications. In Sec. VI,
we present some perspectives for expanding the material that we have
presented here and a conclusion.

\section{The hybrid master equation}

The Lindblad equation,

\begin{equation}
\frac{d}{dt}\hat{\rho}_{SB}=-\frac{i}{\hbar}\left[\hat{H},\hat{\rho}_{SB}\right]+\underset{j}{\sum}\left(\hat{L}_{j}^{\left(S\right)}\hat{\rho}_{SB}\hat{L}_{j}^{\left(S\right)\dagger}-\frac{1}{2}\left\{ \hat{L}_{j}^{\left(S\right)\dagger}\hat{L}_{j}^{\left(S\right)},\hat{\rho}_{SB}\right\} \right)\label{lind1}
\end{equation}
where $\hat{\rho}_{SB}$ is the density operator describing the system
and its environment, $\hat{H}$ is the total Hamiltonian and the $\hat{L}_{j}^{\left(S\right)}$
are the Lindblad operators, which act only on the system, is the most
general form for a master equation \cite{key-7,key-34}. The second
term on the right-hand side of Eq. (\ref{lind1}),
\[
\underset{j}{\sum}\left(\hat{L}_{j}^{\left(S\right)}\hat{\rho}_{SB}\hat{L}_{j}^{\left(S\right)\dagger}-\frac{1}{2}\left\{ \hat{L}_{j}^{\left(S\right)\dagger}\hat{L}_{j}^{\left(S\right)},\hat{\rho}_{SB}\right\} \right),
\]
is the Lindbladian operator acting on the density operator $\hat{\rho}_{SB}.$
The Liouvillian operator acting on $\hat{\rho}_{SB},$
\[
-\frac{i}{\hbar}\left[\hat{H},\hat{\rho}_{SB}\right],
\]
accounts for the unitary portion of the propagation, before the environmental
degrees of freedom are traced out, and the Lindbladian represents
the Markovian measurement dynamics.

We begin with a system $S$ and its environment $B,$ whose interaction
is described by the Hamiltonian:

\begin{eqnarray}
\hat{H}_{SB} & = & \sum_{k}\hat{S}_{k}\hat{B}_{k},\label{defHSB}
\end{eqnarray}
where, for each index $k$, $\hat{S}_{k}$ operates only on the system
$S$ and $\hat{B}_{k},$ only on the environment $B$. The form of
the interaction, Eq. (\ref{defHSB}), is general enough, assumed by
both \emph{amplitude} and \emph{phase damping models} \cite{key-9}.
Here, to account for \emph{non-Markovian} noise, we choose to treat
the environmental interaction as part of the total Hamiltonian, $\hat{H},$
appearing in the Liouvillian term of Eq. (\ref{lind1}). Thus, we
write

\[
\hat{H}=\hat{H}_{B}+\hat{H}_{SB}+\hat{H}_{S},
\]
where $\hat{H}_{B}$ is the environmental Hamiltonian and $\hat{H}_{S}$
is the system Hamiltonian. All the information read by the measuring
apparatus, which is assumed to be \emph{Markovian}, will be accounted
for by the Lindbladian term of Eq. (\ref{lind1}) and, as usual, the
Lindblads $\hat{L}_{j}^{\left(S\right)}$ will act only on the Hilbert
space of the system, since we are interested in measuring system observables
only. Our aim is to obtain an equation for the time evolution of the
reduced density matrix of the system, $\hat{\rho}_{S}$, that is,

\[
\hat{\rho}_{S}=\mathrm{Tr}_{B}\left\{ \hat{\rho}_{SB}\right\} .
\]
For any density matrix operator $\hat{X}$, let the superoperators
$\hat{\hat{B}},$ $\hat{\hat{S}},$ and $\hat{\hat{F}}$ be defined,
respectively, as 
\begin{eqnarray}
\hat{\hat{B}}\hat{X} & = & -\frac{i}{\hbar}\left[\hat{H}_{B},\hat{X}\right],\label{defB}
\end{eqnarray}
\begin{eqnarray*}
\hat{\hat{S}}\hat{X} & = & -\frac{i}{\hbar}\left[\hat{H}_{S},\hat{X}\right]+\underset{j}{\sum}\left(\hat{L}_{j}^{\left(S\right)}\hat{X}\hat{L}_{j}^{\left(S\right)\dagger}-\frac{1}{2}\left\{ \hat{L}_{j}^{\left(S\right)\dagger}\hat{L}_{j}^{\left(S\right)},\hat{X}\right\} \right),
\end{eqnarray*}
and
\begin{eqnarray}
\hat{\hat{F}}\hat{X} & = & -\frac{i}{\hbar}\left[\hat{H}_{SB},\hat{X}\right].\label{defF}
\end{eqnarray}
It is easy to show that $\hat{\hat{B}}\hat{\hat{S}}=\hat{\hat{S}}\hat{\hat{B}}$
and $e^{\hat{\hat{S}}t+\hat{\hat{B}}t}=e^{\hat{\hat{S}}t}e^{\hat{\hat{B}}t}=e^{\hat{\hat{B}}t}e^{\hat{\hat{S}}t}$.

Next, we will also use the Nakajima-Zwanzig projector superoperators
\cite{key-11,key-12}. The defining action of the Nakajima-Zwanzig
projector $\hat{\hat{P}}$ is written as
\begin{eqnarray}
\hat{\hat{P}}\hat{X}\left(t\right) & = & \hat{\rho}_{B}\left(t_{0}\right)\otimes\mathrm{Tr}_{B}\left\{ \hat{X}\left(t\right)\right\} \label{defP}
\end{eqnarray}
for any $\hat{X}\left(t\right)$ and any initial time $t_{0}$. With
these projectors and algebraic manipulations, we obtain the general
equation,
\begin{eqnarray}
\frac{d}{dt}\left[\hat{\hat{P}}\hat{\alpha}\left(t\right)\right] & = & \int_{0}^{t}dt^{\prime}\,\left[\hat{\hat{P}}\hat{\hat{G}}\left(t\right)\hat{\hat{G}}\left(t^{\prime}\right)\hat{\hat{P}}\hat{\alpha}\left(t\right)\right],\label{eqFINAL}
\end{eqnarray}
where 
\begin{eqnarray}
\hat{\alpha}\left(t\right) & = & e^{-\hat{\hat{S}}t-\hat{\hat{B}}t}\hat{\rho}_{SB}\left(t\right)\label{defalfa}
\end{eqnarray}
and
\begin{eqnarray}
\hat{\hat{G}}\left(t\right) & = & e^{-\hat{\hat{S}}t-\hat{\hat{B}}t}\hat{\hat{F}}e^{\hat{\hat{S}}t+\hat{\hat{B}}t}.\label{defG}
\end{eqnarray}
Evidently, according to Eq. (\ref{defalfa}), once $\hat{\alpha}\left(t\right)$
is calculated, $\hat{\rho}_{S}\left(t\right)$ can be found by the
action of $e^{\hat{\hat{S}}t}$ on the reduced $\hat{\alpha}\left(t\right),$
that is, 
\begin{eqnarray*}
\hat{\rho}_{S}\left(t\right) & = & e^{\hat{\hat{S}}t}\mathrm{Tr}_{B}\left\{ \hat{\alpha}\left(t\right)\right\} .
\end{eqnarray*}

As we explain in the Introduction, here we consider a two-state system
interacting with a bath of harmonic oscillators, that emulates a phase-noisy
environment. Thus, we take
\[
\hat{H}_{S}=\hbar\omega_{0}\hat{\sigma}_{z},
\]
\begin{equation}
\hat{H}_{B}=\hbar\underset{k}{\sum}\omega_{k}\hat{b}_{k}^{\dagger}\hat{b}_{k},\label{Hb}
\end{equation}
and \emph{phase-damping }interaction\emph{ }\cite{key-9}, that is,
\[
\begin{cases}
\hat{S}_{k} & =\hbar\hat{\sigma}_{z},\\
\hat{B}_{k} & =g_{k}\hat{b}_{k}^{\dagger}+g_{k}^{*}\hat{b}_{k}.
\end{cases}
\]
where $\omega_{0}$ and the $\omega_{k}$ are real constants, $\hat{b}_{k}$
and $\hat{b}_{k}^{\dagger}$ are the annihilation and creation bath
operators, $g_{k}$ are complex coefficients. Here and next, we will
use the Pauli matrices $\hat{\sigma}_{z}$ and $\hat{\sigma}_{x}$,

\[
\hat{\sigma}_{z}=\left(\begin{array}{cc}
1 & 0\\
0 & -1
\end{array}\right),\:\hat{\sigma}_{x}=\left(\begin{array}{cc}
0 & 1\\
1 & 0
\end{array}\right).
\]

Hence, let us define the operator

\begin{equation}
\hat{B}\equiv\underset{k}{\sum}\hat{B}_{k}=\underset{k}{\sum}\left(g_{k}\hat{b}_{k}^{\dagger}+g_{k}^{*}\hat{b}_{k}\right).\label{opbanho}
\end{equation}
Therefore, the interaction can be written in the simplified form:

\[
\hat{H}_{SB}=\hbar\hat{\sigma}_{z}\hat{B},
\]
that is,
\begin{eqnarray*}
\hat{H}_{SB} & = & \hbar\underset{k}{\sum}\hat{\sigma}_{z}\left(g_{k}\hat{b}_{k}^{\dagger}+g_{k}^{*}\hat{b}_{k}\right).
\end{eqnarray*}
In the case of a finite temperature, we take the initial state of
the environment as given by

\begin{equation}
\hat{\rho}_{B}=\frac{1}{Z_{B}}\underset{k}{\prod}e^{-\hbar\beta\omega_{k}\hat{b}_{k}^{\dagger}\hat{b}_{k}},\, Z_{B}=\underset{l}{\prod}\frac{1}{1-e^{-\hbar\beta\omega_{l}}}.\label{rBtermico}
\end{equation}

\section{Solution of the hybrid master equation}

In this section, we solve Eq. (\ref{eqFINAL}) analytically in the
cases where $\hat{L}^{\left(S\right)}=\lambda\hat{\sigma}_{z}$ at
finite temperature, and $\hat{L}^{\left(S\right)}=\lambda\hat{\sigma}_{x},$
under the assumptions that $\hat{H}_{S}=0$ and $T=0.$ The quantity
$\hat{\hat{P}}\hat{\alpha}\left(t\right)$ appears on both sides of
Eq. (\ref{eqFINAL}) and we can simplify it:

\begin{eqnarray*}
\hat{\hat{P}}\hat{\alpha}\left(t\right) & = & \hat{\hat{P}}e^{-\hat{\hat{S}}t-\hat{\hat{B}}t}\hat{\rho}_{SB}\left(t\right)=e^{-\hat{\hat{S}}t}\hat{\rho}_{S}\left(t\right)\mathrm{Tr}_{B}\left\{ e^{-\hat{\hat{B}}t}\hat{\rho}_{B}\right\} \hat{\rho}_{B}=e^{-\hat{\hat{S}}t}\hat{\rho}_{S}\left(t\right)\hat{\rho}_{B}.
\end{eqnarray*}
Now let us define the operator:

\begin{equation}
\hat{R}\left(t\right)\equiv e^{-\hat{\hat{S}}t}\hat{\rho}_{S}\left(t\right).\label{defR}
\end{equation}
Hence,

\[
\hat{\hat{P}}\hat{\alpha}\left(t\right)=\hat{R}\left(t\right)\hat{\rho}_{B}.
\]
Therefore, to recover the reduced density operator of the system,
we apply $e^{\hat{\hat{S}}t}$ to $\hat{R}\left(t\right):$

\begin{equation}
\hat{\rho}_{S}\left(t\right)=e^{\hat{\hat{S}}t}\hat{R}\left(t\right).\label{roinv}
\end{equation}

An unusual aspect that should be clarified is the action of the superoperator
exponentials, $e^{\hat{\hat{S}}t}$ and $e^{\hat{\hat{B}}t}$. Let
us consider, initially, the time-independent operators $\hat{X}'$
and $\hat{X},$ related by the operation $e^{\hat{\hat{B}}t}$:

\begin{equation}
\hat{X}'=e^{\hat{\hat{B}}t}\hat{X}.\label{aux10}
\end{equation}
When we take the time derivative of $\hat{X}'$ and use Eq. (\ref{aux10}),
we obtain

\[
\frac{d}{dt}\hat{X}'=\hat{\hat{B}}e^{\hat{\hat{B}}t}\hat{X}=\hat{\hat{B}}\hat{X}'.
\]
Now, if we consider the definition of the superoperator $\hat{\hat{B}},$
Eq. (\ref{defB}), we obtain the elementary Liouville-von Neumann
equation, that is,
\begin{eqnarray*}
\frac{d}{dt}\hat{X}' & = & -\frac{i}{\hbar}\left[\hat{H}_{B},\hat{X}'\right],
\end{eqnarray*}
whose solution is easily determined for a time-independent $\hat{H}_{B},$
as in the case of (\ref{Hb}):

\begin{equation}
\hat{X}'=e^{\hat{\hat{B}}t}\hat{X}=e^{-i\frac{\hat{H}_{B}}{\hbar}t}\hat{X}e^{i\frac{\hat{H}_{B}}{\hbar}t}.\label{solsupexpB}
\end{equation}

\subsection{Expanding the integrand that appears in the hybrid master equation}

In view of Eq. (\ref{defR}), the integrand can be written as $\hat{\hat{P}}\hat{\hat{G}}\left(t\right)\hat{\hat{G}}\left(t^{\prime}\right)\hat{R}\left(t\right)\hat{\rho}_{B}$.
From Eq. (\ref{defG}), we obtain 
\begin{eqnarray}
\hat{\hat{G}}\left(t\right)\hat{\hat{G}}\left(t^{\prime}\right)\hat{R}\left(t\right)\hat{\rho}_{B} & = & ie^{-\hat{\hat{S}}t}e^{-\hat{\hat{B}}t}\hat{\hat{F}}\left\{ e^{\hat{\hat{S}}\left(t-t'\right)}\left[\left(e^{\hat{\hat{S}}t'}\hat{R}\left(t\right)\right)\hat{\sigma}_{z}\right]\right\} \nonumber \\
 & \times & \left\{ e^{\hat{\hat{B}}\left(t-t'\right)}\left[\left(e^{\hat{\hat{B}}t'}\hat{\rho}_{B}\right)\hat{B}\right]\right\} \nonumber \\
\nonumber \\
 & - & ie^{-\hat{\hat{S}}t}e^{-\hat{\hat{B}}t}\hat{\hat{F}}\left\{ e^{\hat{\hat{S}}\left(t-t'\right)}\left[\hat{\sigma}_{z}\left(e^{\hat{\hat{S}}t'}\hat{R}\left(t\right)\right)\right]\right\} \nonumber \\
 & \times & \left\{ e^{\hat{\hat{B}}\left(t-t'\right)}\left[\hat{B}\left(e^{\hat{\hat{B}}t'}\hat{\rho}_{B}\right)\right]\right\} .\label{aux14}
\end{eqnarray}
From Eqs. (\ref{defF}) and (\ref{defP}), we can rewrite Eq. (\ref{aux14})
as
\begin{eqnarray}
\hat{\hat{G}}\left(t\right)\hat{\hat{G}}\left(t^{\prime}\right)\hat{R}\left(t\right)\hat{\rho}_{B} & = & e^{-\hat{\hat{S}}t}\hat{\sigma}_{z}\left\{ e^{\hat{\hat{S}}\left(t-t'\right)}\left[\left(e^{\hat{\hat{S}}t'}\hat{R}\left(t\right)\right)\hat{\sigma}_{z}\right]\right\} \nonumber \\
 & \times & \underset{\left(I\right)}{\underbrace{\mathrm{Tr}_{B}\left\{ e^{-\hat{\hat{B}}t}\hat{B}\left\{ e^{\hat{\hat{B}}\left(t-t'\right)}\left[\left(e^{\hat{\hat{B}}t'}\hat{\rho}_{B}\right)\hat{B}\right]\right\} \right\} }}\otimes\hat{\rho}_{B}\nonumber \\
\nonumber \\
 & - & e^{-\hat{\hat{S}}t}\left\{ e^{\hat{\hat{S}}\left(t-t'\right)}\left[\left(e^{\hat{\hat{S}}t'}\hat{R}\left(t\right)\right)\hat{\sigma}_{z}\right]\right\} \hat{\sigma}_{z}\nonumber \\
 & \times & \underset{\left(II\right)}{\underbrace{\mathrm{Tr}_{B}\left\{ e^{-\hat{\hat{B}}t}\left\{ e^{\hat{\hat{B}}\left(t-t'\right)}\left[\left(e^{\hat{\hat{B}}t'}\hat{\rho}_{B}\right)\hat{B}\right]\right\} \hat{B}\right\} }}\otimes\hat{\rho}_{B}\nonumber \\
\nonumber \\
 & - & e^{-\hat{\hat{S}}t}\hat{\sigma}_{z}\left\{ e^{\hat{\hat{S}}\left(t-t'\right)}\left[\hat{\sigma}_{z}\left(e^{\hat{\hat{S}}t'}\hat{R}\left(t\right)\right)\right]\right\} \nonumber \\
 & \times & \underset{\left(III\right)}{\underbrace{\mathrm{Tr}_{B}\left\{ e^{-\hat{\hat{B}}t}\hat{B}\left\{ e^{\hat{\hat{B}}\left(t-t'\right)}\left[\hat{B}\left(e^{\hat{\hat{B}}t'}\hat{\rho}_{B}\right)\right]\right\} \right\} }}\otimes\hat{\rho}_{B}\nonumber \\
\nonumber \\
 & + & e^{-\hat{\hat{S}}t}\left\{ e^{\hat{\hat{S}}\left(t-t'\right)}\left[\hat{\sigma}_{z}\left(e^{\hat{\hat{S}}t'}\hat{R}\left(t\right)\right)\right]\right\} \hat{\sigma}_{z}\nonumber \\
 & \times & \underset{\left(IV\right)}{\underbrace{\mathrm{Tr}_{B}\left\{ e^{-\hat{\hat{B}}t}\left\{ e^{\hat{\hat{B}}\left(t-t'\right)}\left[\hat{B}\left(e^{\hat{\hat{B}}t'}\hat{\rho}_{B}\right)\right]\right\} \hat{B}\right\} }}\otimes\hat{\rho}_{B}.\label{intsep}
\end{eqnarray}
It is interesting to note that, in Eq. (\ref{intsep}), the actions
of $S$ and $B$ are completely separated in each term appearing on
the right-hand side. This proves extremely valuable in the calculations
that follow.

\subsection{Tracing out the environmental degrees of freedom}

For the sake of convenience, let us analyze, firstly, the environment
quantities appearing in Eq. (\ref{intsep}). According to Appendix
A, the partial trace over the environmental variables gives:

\begin{eqnarray}
\hat{\hat{P}}\hat{\hat{G}}\left(t\right)\hat{\hat{G}}\left(t'\right)\hat{\hat{P}}\hat{\alpha}\left(t\right) & = & \left\{ e^{-\hat{\hat{S}}t}\hat{\sigma}_{z}\left\{ e^{\hat{\hat{S}}\left(t-t'\right)}\left[\left(e^{\hat{\hat{S}}t'}\hat{R}\left(t\right)\right)\hat{\sigma}_{z}\right]\right\} \right.\nonumber \\
 &  & \left.-e^{-\hat{\hat{S}}t}\left\{ e^{\hat{\hat{S}}\left(t-t'\right)}\left[\left(e^{\hat{\hat{S}}t'}\hat{R}\left(t\right)\right)\hat{\sigma}_{z}\right]\right\} \hat{\sigma}_{z}\right\} \otimes\hat{\rho}_{B}\nonumber \\
 & \times & \underset{l}{\sum}\left|g_{l}\right|^{2}\left\{ \coth\left(\frac{\hbar\beta\omega_{l}}{2}\right)\cos\left[\omega_{l}\left(t-t'\right)\right]+i\sin\left[\omega_{l}\left(t-t'\right)\right]\right\} \nonumber \\
\nonumber \\
 & + & \left\{ e^{-\hat{\hat{S}}t}\left\{ e^{\hat{\hat{S}}\left(t-t'\right)}\left[\hat{\sigma}_{z}\left(e^{\hat{\hat{S}}t'}\hat{R}\left(t\right)\right)\right]\right\} \hat{\sigma}_{z}\right.\nonumber \\
 &  & \left.-e^{-\hat{\hat{S}}t}\hat{\sigma}_{z}\left\{ e^{\hat{\hat{S}}\left(t-t'\right)}\left[\hat{\sigma}_{z}\left(e^{\hat{\hat{S}}t'}\hat{R}\left(t\right)\right)\right]\right\} \right\} \otimes\hat{\rho}_{B}\nonumber \\
 & \times & \underset{l}{\sum}\left|g_{l}\right|^{2}\left\{ \coth\left(\frac{\hbar\beta\omega_{l}}{2}\right)\cos\left[\omega_{l}\left(t-t'\right)\right]-i\sin\left[\omega_{l}\left(t-t'\right)\right]\right\} .\label{ambfinal}
\end{eqnarray}

\subsection{Introducing a continuous density of states characterizing the environment}

In Eq. (\ref{ambfinal}), if we adopt the general definition of the
density of states as

\begin{equation}
J\left(\omega\right)=\underset{l}{\sum}\left|g_{l}\right|^{2}\delta\left(\omega-\omega_{l}\right),\label{eq:Spectraldensity}
\end{equation}
then the sum over the index $l$ can be replaced by an integral over
a continuum of frequencies:
\begin{eqnarray}
\hat{\hat{P}}\hat{\hat{G}}\left(t\right)\hat{\hat{G}}\left(t'\right)\hat{\hat{P}}\hat{\alpha}\left(t\right) & = & \int_{0}^{\infty}d\omega J\left(\omega\right)\left\{ \coth\left(\frac{\hbar\beta\omega}{2}\right)\cos\left[\omega\left(t-t'\right)\right]+i\sin\left[\omega\left(t-t'\right)\right]\right\} \otimes\hat{\rho}_{B}\nonumber \\
 & \times & \left\{ \underset{\left(A\right)}{\underbrace{e^{-\hat{\hat{S}}t}\hat{\sigma}_{z}\left\{ e^{\hat{\hat{S}}\left(t-t'\right)}\left[\left(e^{\hat{\hat{S}}t'}\hat{R}\left(t\right)\right)\hat{\sigma}_{z}\right]\right\} }}\right.\nonumber \\
\nonumber \\
 & - & \left.\underset{\left(B\right)}{\underbrace{e^{-\hat{\hat{S}}t}\left\{ e^{\hat{\hat{S}}\left(t-t'\right)}\left[\left(e^{\hat{\hat{S}}t'}\hat{R}\left(t\right)\right)\hat{\sigma}_{z}\right]\right\} \hat{\sigma}_{z}}}\right\} \nonumber \\
\nonumber \\
 & + & \int_{0}^{\infty}d\omega J\left(\omega\right)\left\{ \coth\left(\frac{\hbar\beta\omega}{2}\right)\cos\left[\omega\left(t-t'\right)\right]-i\sin\left[\omega\left(t-t'\right)\right]\right\} \otimes\hat{\rho}_{B}\nonumber \\
 & \times & \left\{ \underset{\left(C\right)}{\underbrace{e^{-\hat{\hat{S}}t}\left\{ e^{\hat{\hat{S}}\left(t-t'\right)}\left[\hat{\sigma}_{z}\left(e^{\hat{\hat{S}}t'}\hat{R}\left(t\right)\right)\right]\right\} \hat{\sigma}_{z}}}\right.\nonumber \\
\nonumber \\
 & - & \left.\underset{\left(D\right)}{\underbrace{e^{-\hat{\hat{S}}t}\hat{\sigma}_{z}\left\{ e^{\hat{\hat{S}}\left(t-t'\right)}\left[\hat{\sigma}_{z}\left(e^{\hat{\hat{S}}t'}\hat{R}\left(t\right)\right)\right]\right\} }}\right\} .\label{ambcont}
\end{eqnarray}

Here, to obtain an analytical solution, we choose the Ohmic density
of states (\ref{eq:Spectraldensity}):

\begin{equation}
J\left(\omega\right)=\eta\omega e^{-\frac{\omega}{\omega_{c}}},\label{eq:OhmicSD}
\end{equation}
where $\eta,\omega_{c}\geqslant0$ and $\eta$ is the constant that
gives the intensity of the coupling between the system and its environment.

\subsection{Reduced density operator describing the system}

To obtain the action of the operator $e^{\hat{\hat{S}}t}$, it is
necessary to solve Eq. (\ref{lind1}) without the environment. Then,
we take 
\[
\hat{H}=\hat{H}_{S}=\hbar\omega_{0}\hat{\sigma}_{z}
\]
and, in the Lindbladian, $\hat{L}^{\left(S\right)}=\lambda\hat{\sigma}_{z}$
or $\hat{L}^{\left(S\right)}=\lambda\hat{\sigma}_{x}$.

In the case of $\hat{L}^{\left(S\right)}=\lambda\hat{\sigma}_{z}$,
the solution is simple and can be found in Ref. \cite{key-8}:
\begin{equation}
\begin{cases}
\rho_{11}^{\left(z\right)}\left(t\right) & =\rho_{11}^{\left(z\right)}\left(0\right),\\
\rho_{12}^{\left(z\right)}\left(t\right) & =\rho_{12}^{\left(z\right)}\left(0\right)e^{-2\lambda^{2}t}e^{-i2\omega_{0}t},
\end{cases}\label{zfinal}
\end{equation}
where the upper index $\left(z\right)$ indicates that the solution
is written in the eigenbasis of $\hat{\sigma}_{z}.$

For $\hat{L}^{\left(S\right)}=\lambda\hat{\sigma}_{x}$, the solution
is more complicated:

\begin{equation}
\begin{cases}
\rho_{11}^{\left(z\right)}\left(t\right) & =\frac{1}{2}+\frac{2\rho_{11}^{\left(z\right)}\left(0\right)-1}{2}e^{-2\lambda^{2}t},\\
\\
\rho_{12}^{\left(z\right)}\left(t\right) & =e^{-\lambda^{2}t}\left\{ \rho_{12}^{\left(z\right)}\left(0\right)\cosh\left(\sqrt{\lambda^{4}-4\omega_{0}^{2}}t\right)\right.\\
 & -\rho_{12}^{\left(z\right)}\left(0\right)\frac{i2\omega_{0}}{\sqrt{\lambda^{4}-4\omega_{0}^{2}}}\sinh\left(\sqrt{\lambda^{4}-4\omega_{0}^{2}}t\right)\\
 & +\left.\frac{\lambda^{2}}{\sqrt{\lambda^{4}-4\omega_{0}^{2}}}\rho_{12}^{\left(z\right)*}\left(0\right)\sinh\left(\sqrt{\lambda^{4}-4\omega_{0}^{2}}t\right)\right\} .
\end{cases}\label{solfx}
\end{equation}
To analyze the result of a measurement it is natural to represent
the density operator in the eigenbasis of the measuring apparatus.
Accordingly, in the present case, we use the eigenstates of $\hat{\sigma}_{x},$
that is,

\[
\begin{cases}
\left|+\right\rangle _{x} & =\frac{\left|+\right\rangle +\left|-\right\rangle }{\sqrt{2}},\\
\left|-\right\rangle _{x} & =\frac{\left|+\right\rangle -\left|-\right\rangle }{\sqrt{2}}.
\end{cases}
\]
The change of basis is performed with the eigenvectors matrix 

\begin{equation}
\hat{M}=\frac{1}{\sqrt{2}}\left(\begin{array}{cc}
1 & 1\\
1 & -1
\end{array}\right)=\hat{M}^{-1}\label{M}
\end{equation}
and the result is:
\begin{equation}
\begin{cases}
\rho_{11}^{\left(x\right)}\left(t\right) & =\frac{1}{2}+e^{-\lambda^{2}t}\left\{ \cosh\left(\sqrt{\lambda^{4}-4\omega_{0}^{2}}t\right)\mathrm{Re}\left\{ \rho_{12}^{\left(z\right)}\left(t\right)\right\} \right.\\
 & +\frac{2\omega_{0}}{\sqrt{\lambda^{4}-4\omega_{0}^{2}}}\sinh\left(\sqrt{\lambda^{4}-4\omega_{0}^{2}}t\right)\mathrm{Im}\left\{ \rho_{12}^{\left(z\right)}\left(t\right)\right\} \\
 & +\left.\frac{\lambda^{2}}{\sqrt{\lambda^{4}-4\omega_{0}^{2}}}\sinh\left(\sqrt{\lambda^{4}-4\omega_{0}^{2}}t\right)\mathrm{Re}\left\{ \rho_{12}^{\left(z\right)}\left(t\right)\right\} \right\} ,\\
\\
\rho_{12}^{\left(x\right)}\left(t\right) & =\frac{2\rho_{11}^{\left(z\right)}\left(t\right)-1}{2}e^{-2\lambda^{2}t}-ie^{-\lambda^{2}t}\left\{ \cosh\left(\sqrt{\lambda^{4}-4\omega_{0}^{2}}t\right)\mathrm{Im}\left\{ \rho_{12}^{\left(z\right)}\left(t\right)\right\} \right.\\
 & -\frac{2\omega_{0}}{\sqrt{\lambda^{4}-4\omega_{0}^{2}}}\sinh\left(\sqrt{\lambda^{4}-4\omega_{0}^{2}}t\right)\mathrm{Re}\left\{ \rho_{12}^{\left(z\right)}\left(t\right)\right\} \\
 & -\left.\frac{\lambda^{2}}{\sqrt{\lambda^{4}-4\omega_{0}^{2}}}\sinh\left(\sqrt{\lambda^{4}-4\omega_{0}^{2}}t\right)\mathrm{Im}\left\{ \rho_{12}^{\left(z\right)}\left(t\right)\right\} \right\} .
\end{cases}\label{solfxx}
\end{equation}

\subsection{The case of $\hat{L}^{\left(S\right)}=\lambda\hat{\sigma}_{z}$ }

Here, we consider the evolution of the system in contact with a thermal
reservoir at arbitrary temperature, i.e., we assume that the initial
condition of the environment is given by Eq. (\ref{rBtermico}).

Let us write

\begin{equation}
\hat{R}\left(t\right)=\left(\begin{array}{cc}
R_{11} & R_{12}\\
R_{21} & R_{22}
\end{array}\right),\label{R(t)}
\end{equation}
where, for notational convenience, we take $R_{ij}=R_{ij}\left(t\right).$
Then, using Eq. (\ref{zfinal}), we obtain

\[
e^{\hat{\hat{S}}t'}\hat{R}\left(t\right)=\left(\begin{array}{cc}
R_{11} & R_{12}e^{2\lambda^{2}t'}e^{-i2\omega_{0}t'}\\
R_{21}e^{2\lambda^{2}t'}e^{i2\omega_{0}t'} & R_{22}
\end{array}\right).
\]
Substituting this into Eq. (\ref{ambcont}) and manipulating the terms
according to Appendix B gives

\begin{equation}
\frac{d}{dt}\left(\begin{array}{cc}
R_{11} & R_{12}\\
R_{21} & R_{22}
\end{array}\right)=-4\left(\begin{array}{cc}
0 & R_{12}\\
R_{21} & 0
\end{array}\right)\int_{0}^{t}dt'\int_{0}^{\infty}d\omega J\left(\omega\right)\cos\left[\omega\left(t-t'\right)\right]\coth\left(\frac{\beta\hbar\omega}{2}\right).\label{eqzcont}
\end{equation}

\subsubsection{Obtaining the populations}

According to Eq. (\ref{eqzcont}), the populations are independent
of $J\left(\omega\right)$ and can be immediately evaluated, giving

\[
\frac{d}{dt}R_{ii}=0\Rightarrow R_{ii}\left(t\right)=R_{ii}\left(0\right),
\]
where $i=1,2$. Then, using the constraint that the trace of the density
operator must equal unity, we obtain

\begin{equation}
\begin{cases}
\rho_{11}\left(t\right) & =\rho_{11}\left(0\right),\\
\rho_{22}\left(t\right) & =1-\rho_{11}\left(0\right).
\end{cases}\label{popzfinal}
\end{equation}

\subsubsection{Obtaining the coherences}

From Eq. (\ref{eqzcont}) it follows that the non-diagonal elements
satisfy

\begin{equation}
\frac{d}{dt}R_{ij}=-4\eta R_{ij}\left(t\right)\int_{0}^{t}dt'\int_{0}^{\infty}d\omega\omega e^{-\frac{\omega}{\omega_{c}}}\cos\left[\omega\left(t-t'\right)\right]\coth\left(\frac{\beta\hbar\omega}{2}\right),\label{eqdifcoerz}
\end{equation}
where $i,j=1,2$ and $i\neq j$. From the procedure of Appendix C,
we obtain

\begin{eqnarray}
\rho_{12}\left(t\right) & =\rho_{12}\left(0\right)\left[\frac{\Gamma\left(\frac{1}{\omega_{c}\beta\hbar}+i\frac{t}{\beta\hbar}\right)\Gamma\left(\frac{1}{\omega_{c}\beta\hbar}-i\frac{t}{\beta\hbar}\right)}{\Gamma^{2}\left(\frac{1}{\omega_{c}\beta\hbar}\right)}\frac{\Gamma\left(\frac{1}{\omega_{c}\beta\hbar}+1+i\frac{t}{\beta\hbar}\right)\Gamma\left(\frac{1}{\omega_{c}\beta\hbar}+1-i\frac{t}{\beta\hbar}\right)}{\Gamma^{2}\left(\frac{1}{\omega_{c}\beta\hbar}+1\right)}\right]^{2\eta} & e^{-2\lambda^{2}t}e^{i2\omega_{0}t},\label{ro12z}
\end{eqnarray}
where $\Gamma$ denotes the gamma function. Simplifying Eq. (\ref{ro12z}),
according to Appendix D, gives

\begin{equation}
\rho_{12}\left(t\right)=\rho_{12}\left(0\right)\left\{ \frac{1}{1+\left(\omega_{c}t\right)^{2}}\left[\underset{n=1}{\overset{\infty}{\prod}}\frac{1+n\omega_{c}\beta\hbar}{1+n\omega_{c}\beta\hbar+\omega_{c}t}\right]^{2}\right\} ^{2\eta}e^{-2\lambda^{2}t}e^{i2\omega_{0}t}.\label{r12zfinal}
\end{equation}

The consistency of Eqs. (\ref{ro12z}) and (\ref{r12zfinal}) with
the case without environment can be verified by noticing that, in
the limit $\eta\rightarrow0$ (i.e., in the absence of environment),
they reduce to Eq. (\ref{zfinal}), as expected.

\subsection{The case of $\hat{L}^{\left(S\right)}=\lambda\hat{\sigma}_{x}$}

From Eq. (\ref{solfx}) the action of $e^{\hat{\hat{S}}t}$ on an
operator
\begin{eqnarray*}
\hat{x} & = & \left(\begin{array}{cc}
x_{11} & x_{12}\\
x_{21} & x_{22}
\end{array}\right)
\end{eqnarray*}
gives

\[
e^{\hat{\hat{S}}t}\hat{x}=\left(\begin{array}{cc}
s_{11} & s_{12}\\
s_{21} & s_{22}
\end{array}\right),
\]
where

\begin{equation}
\begin{cases}
s_{11}= & \frac{x_{11}-x_{22}}{2}e^{-2\lambda^{2}t}+\frac{x_{11}+x_{22}}{2},\\
s_{22}= & -\frac{x_{11}-x_{22}}{2}e^{-2\lambda^{2}t}+\frac{x_{11}+x_{22}}{2},
\end{cases}\label{aux29}
\end{equation}

\begin{equation}
\begin{cases}
s_{12}= & \frac{e^{-\lambda^{2}t}}{\Omega}\left[\Omega\cosh\left(\Omega t\right)x_{12}-i2\omega_{0}\sinh\left(\Omega t\right)x_{12}+\lambda^{2}\sinh\left(\Omega t\right)x_{21}\right],\\
s_{21}= & \frac{e^{-\lambda^{2}t}}{\Omega}\left[\Omega\cosh\left(\Omega t\right)x_{21}+i2\omega_{0}\sinh\left(\Omega t\right)x_{21}+\lambda^{2}\sinh\left(\Omega t\right)x_{12}\right],
\end{cases}\label{aux30}
\end{equation}
with

\[
\Omega\equiv\sqrt{\lambda^{4}-4\omega_{0}^{2}}.
\]

According to the procedure explained in Appendix E, using Eqs. (\ref{aux29})
and (\ref{aux30}) in the case of the operator

\begin{equation}
\hat{R}\left(t\right)=\left(\begin{array}{cc}
r_{11}\left(t\right) & r_{12}\left(t\right)\\
r_{21}\left(t\right) & r_{22}\left(t\right)
\end{array}\right)\label{Rx}
\end{equation}
results in the final differential equation:

\begin{eqnarray}
\frac{d}{dt}\left[\begin{array}{cc}
r_{11}\left(t\right) & r_{12}\left(t\right)\\
r_{21}\left(t\right) & r_{22}\left(t\right)
\end{array}\right] & = & -\frac{4}{\Omega^{3}}\int_{0}^{t}dt'\int_{0}^{\infty}d\omega J\left(\omega\right)\cos\left[\omega\left(t-t'\right)\right]\coth\left(\frac{\beta\hbar\omega}{2}\right)\nonumber \\
 & \times & \left[\begin{array}{cc}
0 & Q_{1}\left(t,t'\right)r_{12}\left(t\right)+Q_{2}\left(t,t'\right)r_{21}\left(t\right)\\
Q_{2}^{*}\left(t,t'\right)r_{12}\left(t\right)+Q_{1}^{*}\left(t,t'\right)r_{21}\left(t\right) & 0
\end{array}\right],\label{eqmatfinal}
\end{eqnarray}
where

\begin{equation}
\begin{cases}
Q_{1}\left(t,t'\right)\equiv & K_{1}\left(t\right)\left[K_{1}^{*}\left(t-t'\right)K_{1}^{*}\left(t'\right)-K_{2}\left(t-t'\right)K_{2}\left(t'\right)\right]\\
 & +K_{2}\left(t\right)\left[K_{2}\left(t-t'\right)K_{1}^{*}\left(t'\right)-K_{1}\left(t-t'\right)K_{2}\left(t'\right)\right],\\
Q_{2}\left(t,t'\right)\equiv & K_{1}\left(t\right)\left[K_{1}^{*}\left(t-t'\right)K_{2}\left(t'\right)-K_{2}\left(t-t'\right)K_{1}\left(t'\right)\right]\\
 & +K_{2}\left(t\right)\left[K_{2}\left(t-t'\right)K_{2}\left(t'\right)-K_{1}\left(t-t'\right)K_{1}\left(t'\right)\right],
\end{cases}\label{Q}
\end{equation}
and
\[
\begin{cases}
K_{1}\left(t\right)\equiv & \Omega\cosh\left(\Omega t\right)+i2\omega_{0}\sinh\left(\Omega t\right),\\
K_{2}\left(t\right)\equiv & \lambda^{2}\sinh\left(\Omega t\right).
\end{cases}
\]

\subsubsection{The populations represented in the eigenbasis of $\hat{\sigma}_{z}$}

Analogously to the case of $\hat{L}^{\left(S\right)}=\lambda\hat{\sigma}_{z}$
and $T\neq0$, the populations do not depend on $J\left(\omega\right)$.
From Eq. (\ref{eqmatfinal}) we obtain
\[
\frac{d}{dt}r_{ii}=0\Rightarrow r_{ii}\left(t\right)=r_{ii}\left(0\right),
\]
where $i=1,2$. We notice that $r_{ii}\left(0\right)=\rho_{11}^{\left(z\right)}\left(0\right)$
and, using Eq. (\ref{roinv}), we can write:

\[
\rho_{11}^{\left(z\right)}\left(t\right)=\frac{2\rho_{11}^{\left(z\right)}\left(0\right)-1}{2}e^{-2\lambda^{2}t}+\frac{1}{2}.
\]

\subsubsection{The coherences represented in the eigenbasis of $\hat{\sigma}_{z}$,
for $T=0$ and $\omega_{0}=0$}

Since $r_{12}\left(t\right)$ is the complex conjugate of $r_{21}\left(t\right)$,
we only need to calculate one of them. Let us introduce the new variable
$\tau=t-t'.$ Hence, it follows from Eq. (\ref{eqmatfinal}) that

\begin{eqnarray}
\frac{d}{dt}r_{12}\left(t\right) & = & -4\frac{\eta}{\Omega^{3}}\int_{0}^{t}d\tau\int_{0}^{\infty}d\omega\omega e^{-\frac{\omega'}{\omega_{c}}}\cos\left(\omega\tau\right)\coth\left(\frac{\beta\hbar\omega}{2}\right)\nonumber \\
 & \times & \left[Q_{1}\left(t,t-\tau\right)r_{12}\left(t\right)+Q_{2}\left(t,t-\tau\right)r_{21}\left(t\right)\right].\label{Tnnulo}
\end{eqnarray}
There is no analytic solution for this equation at a finite temperature.
However, as detailed in Appendix F, we have found the following result
for $T=0$ and $\omega_{0}=0:$
\begin{eqnarray}
\rho_{12}^{\left(z\right)}\left(t\right) & = & \mathrm{Re}\left\{ \rho_{12}^{\left(z\right)}\left(0\right)\right\} e^{-8\eta\lambda^{2}g_{0}t}e^{4\eta\lambda^{2}\left[A_{-}\left(t\right)-B_{-}\left(t\right)\right]}\nonumber \\
 & + & i\mathrm{Im}\left\{ \rho_{12}^{\left(z\right)}\left(0\right)\right\} e^{-2\lambda^{2}t}e^{8\eta\lambda^{2}g_{0}t}e^{-4\eta\lambda^{2}\left[A_{+}\left(t\right)+B_{+}\left(t\right)\right]},\label{sxcoeraprox}
\end{eqnarray}
where
\begin{eqnarray*}
A_{+}\left(t\right) & \equiv & \int_{0}^{t}e^{2\lambda^{2}t'}g_{1}\left(t'\right)dt'\\
 & = & \frac{1}{4\lambda^{2}}\mathrm{Re}\left\{ e^{i2\frac{\lambda^{2}}{\omega_{c}}}\left[e^{4\lambda^{2}t}\Gamma\left(0,2\lambda^{2}t+i2\frac{\lambda^{2}}{\omega_{c}}\right)-e^{-i4\frac{\lambda^{2}}{\omega_{c}}}\Gamma\left(0,-2\lambda^{2}t-i2\frac{\lambda^{2}}{\omega_{c}}\right)\right]\right\} \\
 & + & \frac{1}{2\lambda^{2}}\mathrm{Re}\left\{ e^{-i2\frac{\lambda^{2}}{\omega_{c}}}\left[i2\frac{\lambda^{2}}{\omega_{c}}\Gamma\left(0,-2\lambda^{2}t-i2\frac{\lambda^{2}}{\omega_{c}}\right)+\Gamma\left(1,-2\lambda^{2}t-i2\frac{\lambda^{2}}{\omega_{c}}\right)\right]\right\} \\
 & + & t\mathrm{Re}\left\{ e^{-i2\frac{\lambda^{2}}{\omega_{c}}}\Gamma\left(0,-2\lambda^{2}t-i2\frac{\lambda^{2}}{\omega_{c}}\right)\right\} +\mathrm{Re}\left\{ e^{-i2\frac{\lambda^{2}}{\omega_{c}}}c_{1}+e^{i2\frac{\lambda^{2}}{\omega_{c}}}c_{3}\right\} ,
\end{eqnarray*}

\begin{eqnarray*}
A_{-}\left(t\right) & \equiv & \int_{0}^{t}e^{-2\lambda^{2}t'}g_{1}\left(t'\right)dt'\\
 & = & -\frac{1}{4\lambda^{2}}\mathrm{Re}\left\{ e^{-i2\frac{\lambda^{2}}{\omega_{c}}}\left[e^{-4\lambda^{2}t}\Gamma\left(0,-2\lambda^{2}t-i2\frac{\lambda^{2}}{\omega_{c}}\right)-e^{i4\frac{\lambda^{2}}{\omega_{c}}}\Gamma\left(0,2\lambda^{2}t+i2\frac{\lambda^{2}}{\omega_{c}}\right)\right]\right\} \\
 & + & \frac{1}{2\lambda^{2}}\mathrm{Re}\left\{ e^{i2\frac{\lambda^{2}}{\omega_{c}}}\left[i2\frac{\lambda^{2}}{\omega_{c}}\Gamma\left(0,2\lambda^{2}t+i2\frac{\lambda^{2}}{\omega_{c}}\right)-\Gamma\left(1,2\lambda^{2}t+i2\frac{\lambda^{2}}{\omega_{c}}\right)\right]\right\} \\
 & + & t\mathrm{Re}\left\{ e^{i2\frac{\lambda^{2}}{\omega_{c}}}\Gamma\left(0,2\lambda^{2}t+i2\frac{\lambda^{2}}{\omega_{c}}\right)\right\} -\mathrm{Re}\left\{ e^{i2\frac{\lambda^{2}}{\omega_{c}}}c_{1}^{*}+e^{-i2\frac{\lambda^{2}}{\omega_{c}}}c_{3}^{*}\right\} ,
\end{eqnarray*}

\begin{eqnarray*}
B_{+}\left(t\right) & \equiv & \int_{0}^{t}e^{2\lambda^{2}t'}g_{2}\left(t'\right)dt'\\
 & = & \frac{1}{4\lambda^{2}}\mathrm{Re}\left\{ e^{i2\frac{\lambda^{2}}{\omega_{c}}}\left[e^{4\lambda^{2}t}\Gamma\left(-1,2\lambda^{2}t+i2\frac{\lambda^{2}}{\omega_{c}}\right)+e^{-i4\frac{\lambda^{2}}{\omega_{c}}}\Gamma\left(-1,-2\lambda^{2}t-i2\frac{\lambda^{2}}{\omega_{c}}\right)\right]\right\} \\
 & - & \frac{1}{2\lambda^{2}}\mathrm{Re}\left\{ e^{-i2\frac{\lambda^{2}}{\omega_{c}}}\left[i2\frac{\lambda^{2}}{\omega_{c}}\Gamma\left(-1,-2\lambda^{2}t-i2\frac{\lambda^{2}}{\omega_{c}}\right)+\Gamma\left(0,-2\lambda^{2}t-i2\frac{\lambda^{2}}{\omega_{c}}\right)\right]\right\} \\
 & - & t\mathrm{Re}\left\{ e^{-i2\frac{\lambda^{2}}{\omega_{c}}}\Gamma\left(-1,-2\lambda^{2}t-i2\frac{\lambda^{2}}{\omega_{c}}\right)\right\} -\mathrm{Re}\left\{ e^{-i2\frac{\lambda^{2}}{\omega_{c}}}c_{2}-e^{i2\frac{\lambda^{2}}{\omega_{c}}}c_{4}\right\} ,
\end{eqnarray*}

\begin{eqnarray*}
B_{-}\left(t\right) & \equiv & \int_{0}^{t}e^{-2\lambda^{2}t'}g_{2}\left(t'\right)dt'\\
 & = & \frac{1}{4\lambda^{2}}\mathrm{Re}\left\{ e^{-i2\frac{\lambda^{2}}{\omega_{c}}}\left[e^{-4\lambda^{2}t}\Gamma\left(-1,-2\lambda^{2}t-i2\frac{\lambda^{2}}{\omega_{c}}\right)+e^{i4\frac{\lambda^{2}}{\omega_{c}}}\Gamma\left(-1,2\lambda^{2}t+i2\frac{\lambda^{2}}{\omega_{c}}\right)\right]\right\} \\
 & + & \frac{1}{2\lambda^{2}}\mathrm{Re}\left\{ e^{i2\frac{\lambda^{2}}{\omega_{c}}}\left[i2\frac{\lambda^{2}}{\omega_{c}}\Gamma\left(-1,2\lambda^{2}t+i2\frac{\lambda^{2}}{\omega_{c}}\right)-\Gamma\left(0,2\lambda^{2}t+i2\frac{\lambda^{2}}{\omega_{c}}\right)\right]\right\} \\
 & + & t\mathrm{Re}\left\{ e^{i2\frac{\lambda^{2}}{\omega_{c}}}\Gamma\left(-1,2\lambda^{2}t+i2\frac{\lambda^{2}}{\omega_{c}}\right)\right\} -\mathrm{Re}\left\{ e^{i2\frac{\lambda^{2}}{\omega_{c}}}c_{2}^{*}-e^{-i2\frac{\lambda^{2}}{\omega_{c}}}c_{4}^{*}\right\} ,
\end{eqnarray*}
and

\[
\begin{cases}
c_{1}= & -\frac{1}{2\lambda^{2}}\left[i\frac{2\lambda^{2}}{\omega_{C}}\Gamma\left(0,-i\frac{2\lambda^{2}}{\omega_{C}}\right)+\Gamma\left(1,-i\frac{2\lambda^{2}}{\omega_{C}}\right)\right],\\
c_{2}= & -\frac{1}{2\lambda^{2}}\left[i\frac{2\lambda^{2}}{\omega_{C}}\Gamma\left(-1,-i\frac{2\lambda^{2}}{\omega_{C}}\right)+\Gamma\left(0,-i\frac{2\lambda^{2}}{\omega_{C}}\right)\right],\\
c_{3}= & -\frac{1}{4\lambda^{2}}\left[\Gamma\left(0,i\frac{2\lambda^{2}}{\omega_{C}}\right)-e^{-i\frac{4\lambda^{2}}{\omega_{C}}}\Gamma\left(0,-i\frac{2\lambda^{2}}{\omega_{C}}\right)\right],\\
c_{4}= & -\frac{1}{4\lambda^{2}}\left[\Gamma\left(-1,i\frac{2\lambda^{2}}{\omega_{C}}\right)+e^{-i\frac{4\lambda^{2}}{\omega_{C}}}\Gamma\left(-1,-i\frac{2\lambda^{2}}{\omega_{C}}\right)\right].
\end{cases}
\]

\subsubsection{The density matrix elements represented in the eigenbasis of $\hat{\sigma}_{x}$,
for $T=0$ and $\omega_{0}=0$}

Let us change the previous result for the coherences to the eigenbasis
of $\hat{\sigma}_{x}.$ Using Eq. (\ref{solfxx}) with $\omega_{0}=0$
we obtain 
\begin{equation}
\begin{cases}
\hat{\rho}_{11}^{\left(x\right)}\left(t\right)= & \frac{1}{2}+\mathrm{Re}\left\{ \rho_{12}^{\left(z\right)}\left(0\right)\right\} e^{-8\eta\lambda^{2}g_{0}t}e^{4\eta\lambda^{2}\left[A_{-}\left(t\right)-B_{-}\left(t\right)\right]},\\
\hat{\rho}_{12}^{\left(x\right)}\left(t\right)= & \frac{2\rho_{11}^{\left(z\right)}\left(0\right)-1}{2}e^{-2\lambda^{2}t}-i\mathrm{Im}\left\{ \rho_{12}^{\left(z\right)}\left(0\right)\right\} e^{-2\lambda^{2}t}e^{8\eta\lambda^{2}g_{0}t}e^{-4\eta\lambda^{2}\left[A_{+}\left(t\right)+B_{+}\left(t\right)\right]}.
\end{cases}\label{roxfinal}
\end{equation}

\section{Numerical method}

The approximate analytical results detailed above have been compared
against the exact numerical solution of the Lindblad equation, Eq.
(\ref{lind1}). The algorithm we use derives from the superoperator-splitting
method \cite{key-41}, which is adequate for numerically solving a
first-order differential equation of the form:

\[
\frac{d}{dt}\hat{\rho}_{SB}\left(t\right)=\hat{\hat{A}}\hat{\rho}_{SB}\left(t\right)+\hat{\hat{B}}\hat{\rho}_{SB}\left(t\right).
\]
As long as the two superoperators $\hat{\hat{A}}$ and $\hat{\hat{B}}$
are time-independent, the solution for the differential equation is
\begin{eqnarray*}
\hat{\rho}_{SB}\left(t\right) & = & e^{\left(\hat{\hat{A}}+\hat{\hat{B}}\right)t}\hat{\rho}_{SB}\left(0\right).
\end{eqnarray*}
In case this exponential superoperator cannot be analytically found,
but the $e^{\hat{\hat{A}}t}$ and $e^{\hat{\hat{B}}t}$ can, we may
use the approximation
\begin{eqnarray*}
e^{\left(\hat{\hat{A}}+\hat{\hat{B}}\right)\Delta t} & = & e^{\hat{\hat{A}}\Delta t}e^{\hat{\hat{B}}\Delta t}+O\left(\Delta t^{2}\right)
\end{eqnarray*}
to expand the solution in terms of a product of $N$ short time steps
of length $\Delta t$. The alternate application of these two superoperators,
$e^{\hat{\hat{A}}\Delta t}$ and $e^{\hat{\hat{B}}\Delta t},$ comprises
the superoperator-splitting method.

For a finite-time measurement, the superoperator $\hat{\hat{A}}$
represents the Liouvillian and $\hat{\hat{B}}$ represents the Lindbladian.
The form of the exponential of the time-independent Liouvillian is
straightforward, while $e^{\hat{\hat{B}}\Delta t}$ behaves in the
way described by Eqs. (\ref{aux29}) and (\ref{aux30}). As the behavior
of both populations in the $\hat{\sigma}_{z}$ eigenbasis can be exactly
described by a simple analytical formula, only the coherences need
to be calculated numerically. Therefore, all the relevant information
about the density matrix can be represented by a pseudo-spinor containing
both coherences, on which the alternate action of $e^{\hat{\hat{A}}\Delta t}$
and $e^{\hat{\hat{B}}\Delta t}$ will be equivalent to matrix products:

\begin{equation}
\left(\begin{array}{c}
\hat{\rho}_{12}\left(N\Delta t\right)\\
\hat{\rho}_{21}\left(N\Delta t\right)
\end{array}\right)=e^{-\lambda^{2}\left(N\Delta t\right)}\left[\hat{\hat{K}}_{+1}\left(\Delta t\right)A_{+1}\left(\Delta t\right)+\hat{\hat{K}}_{-1}\left(\Delta t\right)A_{-1}\left(\Delta t\right)\right]^{N}\left(\begin{array}{c}
\hat{\rho}_{12}\left(0\right)\\
\hat{\rho}_{21}\left(0\right)
\end{array}\right),\label{eq:algorithm1}
\end{equation}
where
\begin{eqnarray*}
\hat{\hat{K}}_{q}\left(\Delta t\right)\hat{X} & \equiv & e^{-i\sum_{k}\omega_{k}\left(\hat{b}_{k}+qg_{k}/\omega_{k}\right)^{\dagger}\left(\hat{b}_{k}+qg_{k}/\omega_{k}\right)\Delta t}\hat{X}e^{i\sum_{k}\omega_{k}\left(\hat{b}_{k}-qg_{k}/\omega_{k}\right)^{\dagger}\left(\hat{b}_{k}-qg_{k}/\omega_{k}\right)\Delta t}
\end{eqnarray*}
and we defined the square matrices $A_{\pm}\left(\Delta t\right)$
as

\begin{eqnarray*}
A_{+1}\left(\Delta t\right) & \equiv & \left(\begin{array}{cc}
b_{+1}\left(\Delta t\right) & b_{-1}\left(\Delta t\right)\\
0 & 0
\end{array}\right),\\
A_{-1}\left(\Delta t\right) & \equiv & \left(\begin{array}{cc}
0 & 0\\
b_{-1}\left(\Delta t\right) & b_{+1}\left(\Delta t\right)
\end{array}\right),
\end{eqnarray*}
with
\begin{eqnarray*}
b_{q}\left(\Delta t\right) & \equiv & \frac{1}{2}\left(e^{\lambda^{2}\Delta t}+qe^{-\lambda^{2}\Delta t}\right).
\end{eqnarray*}

The binomial in Eq. (\ref{eq:algorithm1}) may be expanded in $N$
two-term summations. Also, as we are not interested in the total density
matrix, but in the reduced part referring to the two-state system,
we may take the partial trace over the degrees of freedom of the environment,
to find:

\begin{eqnarray}
\left(\begin{array}{c}
\rho_{12}\left(N\Delta t\right)\\
\rho_{21}\left(N\Delta t\right)
\end{array}\right) & = & e^{-\lambda^{2}\left(N\Delta t\right)}\sum_{q_{1}\in\left\{ -1,1\right\} }\ldots\sum_{q_{N}\in\left\{ -1,1\right\} }\prod_{n=1}^{N}\left[A_{q_{n}}\left(\Delta t\right)\right]\nonumber \\
 & \times & \mathrm{Tr}_{B}\left\{ \prod_{n=1}^{N}\left[\hat{\hat{K}}_{q_{n}}\left(\Delta t\right)\right]\left|0\right\rangle \left\langle 0\right|\right\} \left(\begin{array}{c}
\rho_{12}\left(0\right)\\
\rho_{21}\left(0\right)
\end{array}\right),\label{eq:algorithm2}
\end{eqnarray}
where we are considering the system initially separable from the environment,
which starts in the vacuum state ($T=0$).

Employing the result for the trace given in Appendix G and the easily-verified
matrix identity,
\begin{eqnarray*}
A_{r}\left(\Delta t\right)A_{s}\left(\Delta t\right) & = & b_{rs}\left(\Delta t\right)\left(\sigma_{x}\right)^{\left(1-rs\right)/2}A_{s}\left(\Delta t\right),
\end{eqnarray*}
we find the final form for the algorithm, which is comprised of a
sum of $2^{N}$ terms, each containing $N^{2}\left(N-1\right)$ factors
and a product of up to two matrices:

\begin{eqnarray}
\left(\begin{array}{c}
\rho_{12}\left(N\Delta t\right)\\
\rho_{21}\left(N\Delta t\right)
\end{array}\right) & = & e^{-\lambda^{2}\left(N\Delta t\right)}\sum_{q_{1}\in\left\{ -1,1\right\} }\ldots\sum_{q_{N}\in\left\{ -1,1\right\} }\nonumber \\
 &  & \prod_{m,n=1}^{N}\left\{ 1+\frac{2\left(\omega_{c}\Delta t\right)^{-2}+\left(1-2\left|m-n\right|^{2}\right)}{\left[\left(\omega_{c}\Delta t\right)^{-2}+\left|m-n\right|^{2}\right]^{2}}\right\} ^{-\eta q_{m}q_{n}}\nonumber \\
 & \times & \prod_{n=1}^{N-1}\left[b_{q_{n}q_{n+1}}\left(\Delta t\right)\right]\left(\sigma_{x}\right)^{\left(1-q_{1}q_{N}\right)/2}A_{q_{N}}\left(\Delta t\right)\left(\begin{array}{c}
\rho_{12}\left(0\right)\\
\rho_{21}\left(0\right)
\end{array}\right),\label{eq:algorithm3}
\end{eqnarray}
which can be implemented in any programming language capable of handling
large floating-point numbers and of calculating powers and exponentials.
The results are then translated into elements of the density matrix
in the basis of eigenvectors of $\hat{\sigma}_{x}$ by a simple transformation,
Eq. (\ref{M}).

\section{Results and Discussion}

The final analytical expression for the finite-time $\hat{\sigma}_{x}$
measurement given in Eq. (\ref{roxfinal}) is expected to agree with
the numerical solution Eq. (\ref{eq:algorithm3}) for weak coupling
with the environment, in our case represented by the dimensionless
constant $\eta$. This fact is confirmed in Fig. 1, which compares
the time evolution of the population $\rho_{11}\left(t\right)$ using
both methods for different values of the system-environment coupling
constant. There it can be seen that the discrepancy between the two
graphs grows with $\eta$.

\begin{figure}
\includegraphics{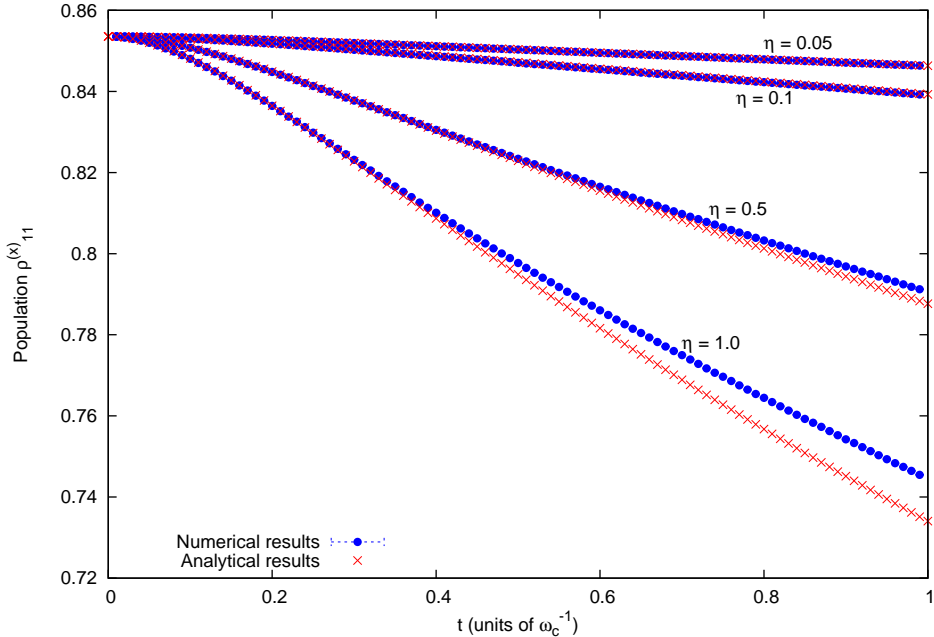}\caption{(Color online) Comparison between the time evolution of the population
in the eigenbasis of $\hat{\sigma}_{x}$, $\rho_{11}^{\left(x\right)}\left(t\right)$
, given by the analytical (crosses) and numerical (bullets) methods
for $\lambda^{2}=4\omega_{c}$ and different values of $\eta$. The
system is initially in the pure state $\frac{1}{\sqrt{2}}\left(\left|1\right\rangle +e^{i\pi/4}\left|2\right\rangle \right)$.}
\end{figure}

Having confirmed the reliability of our results, we proceed to compare
the noisy measurement described by our formalism with the case where
the system is not being measured. As can be seen in Fig. 2, systems
evolving only under the influence of the environment ($\lambda^{2}=0$)
always have faster rates of change of their populations than those
which are also interacting with the measurement apparatus ($\lambda^{2}>0$).
Moreover, the stronger the measurement, the slower the rate of change
of the population. Therefore, a measurement of the observable $\hat{\sigma}_{x}$
helps to maintain the population closer to its original value.

\begin{figure}
\includegraphics{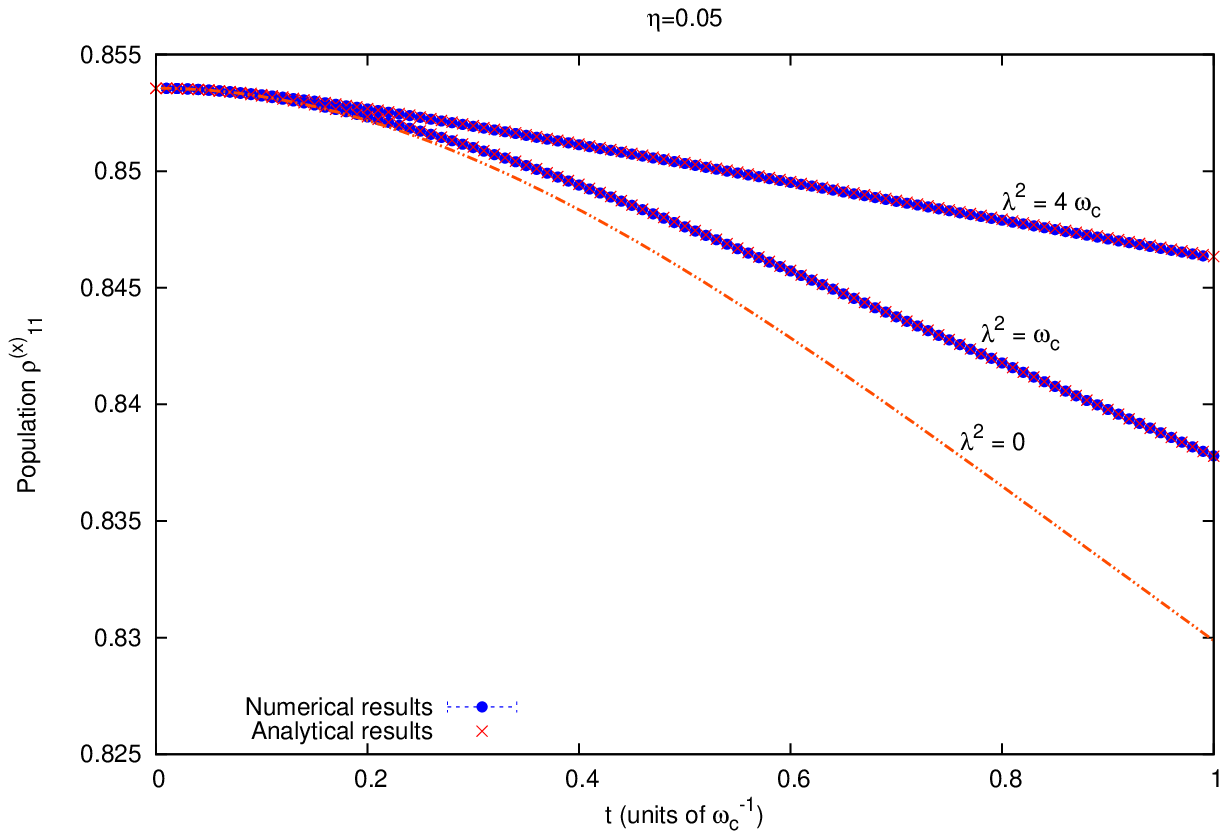}

\includegraphics{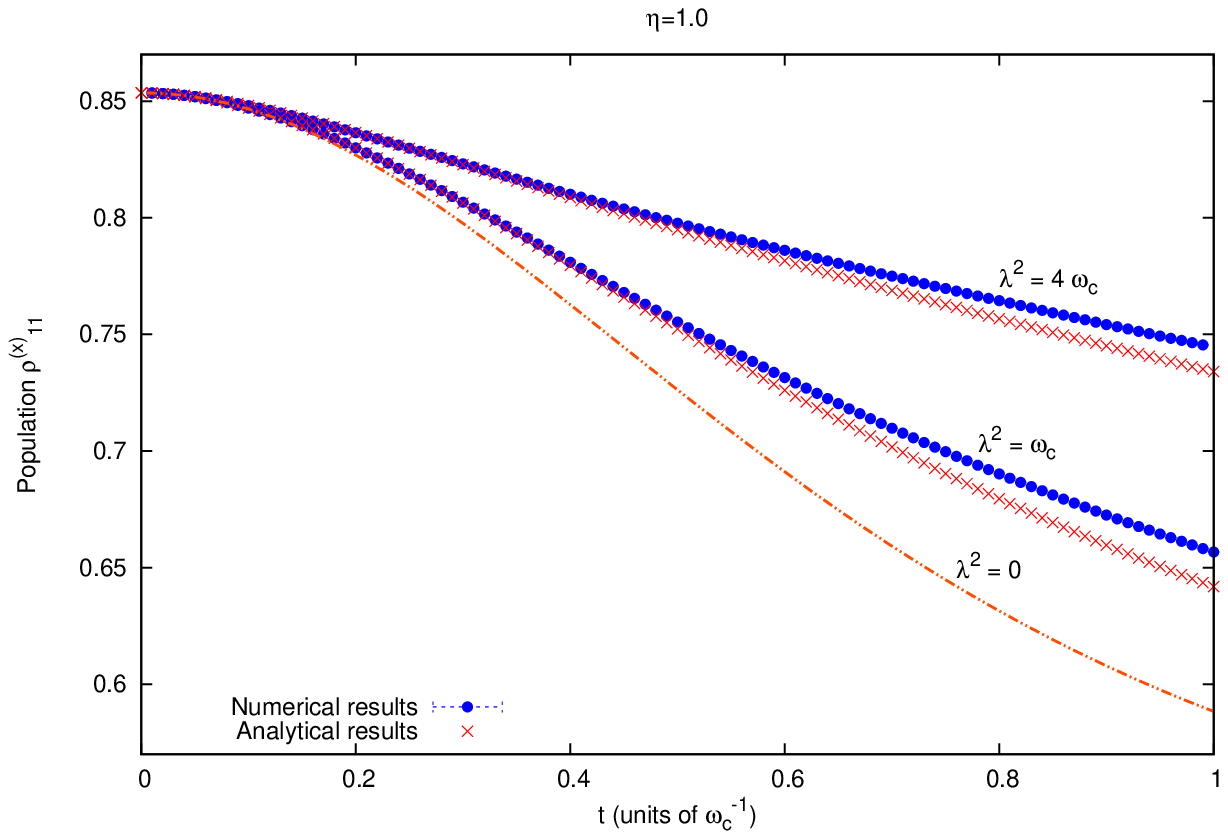}\caption{(Color online) Time evolution of the population in the eigenbasis
of $\hat{\sigma}_{x}$ when there is an external environment causing
phase noise (crosses for the analytical method and bullets for the
numerical). The dashed line represents the evolution of a system that
is only subject to the interaction with the environment ($\lambda=0$).
In all graphs, the evolution of the system that is not subject to
measurement diverges faster from the initial value than those systems
that are being measured. The initial conditions are the same applied
to Fig. 1.}
\end{figure}

The meaning of this phenomenon is straightforward: a difference between
the initial value of the population (instant $t_{0}$) and its value
at the end of the measurement (instant $t_{f}$) represents the probability
that the measurement will give a wrong result. This is so because,
if the initial value $\rho_{11}\left(t_{0}\right)$ and the final
result $\rho_{11}\left(t_{f}\right)$ are different, then there will
be a fraction of $\epsilon=\left|\rho_{11}\left(t_{f}\right)-\rho_{11}\left(t_{0}\right)\right|$
systems in the statistical ensemble that present final collapsed states
different from those that would be obtained if we had measured them
with an ideal instantaneous process at instant $t_{0}$. We shall
henceforth refer to the difference $\epsilon$ as the error.

The fact that a finite-time measurement helps to preserve the initial
value of the population of a system also shows that a naïve approach
to a noisy measurement will overestimate the error. Given that a measurement
ends at a time $t_{f}$ and that the system has been interacting with
the environment since instant $t_{0}<t_{f}$, modeling the noisy measurement
process as a period of interaction with the environment followed by
an instantaneous measurement would lead to an error much larger than
that obtained by considering that the system is measured continuously
during a period $t_{f}-t_{0}$. Therefore, a description of the measurement
as a continuous process is essential to have good estimates of the
error.

Finally, it is important to notice that this protection against error
depends on whether the observable measured commutes with the interaction
Hamiltonian. In the measurement described above, the observable $\hat{\sigma}_{x}$
anti-commutes with the operator $\hat{\sigma}_{z}$ in the Hamiltonian.
But when we measure an observable that does commute, the populations
remain constant, and the only effect of the measuring apparatus is
to strengthen the effects of the decoherence caused by the environment,
as can be seen from Fig. 3.

\begin{figure}
\includegraphics{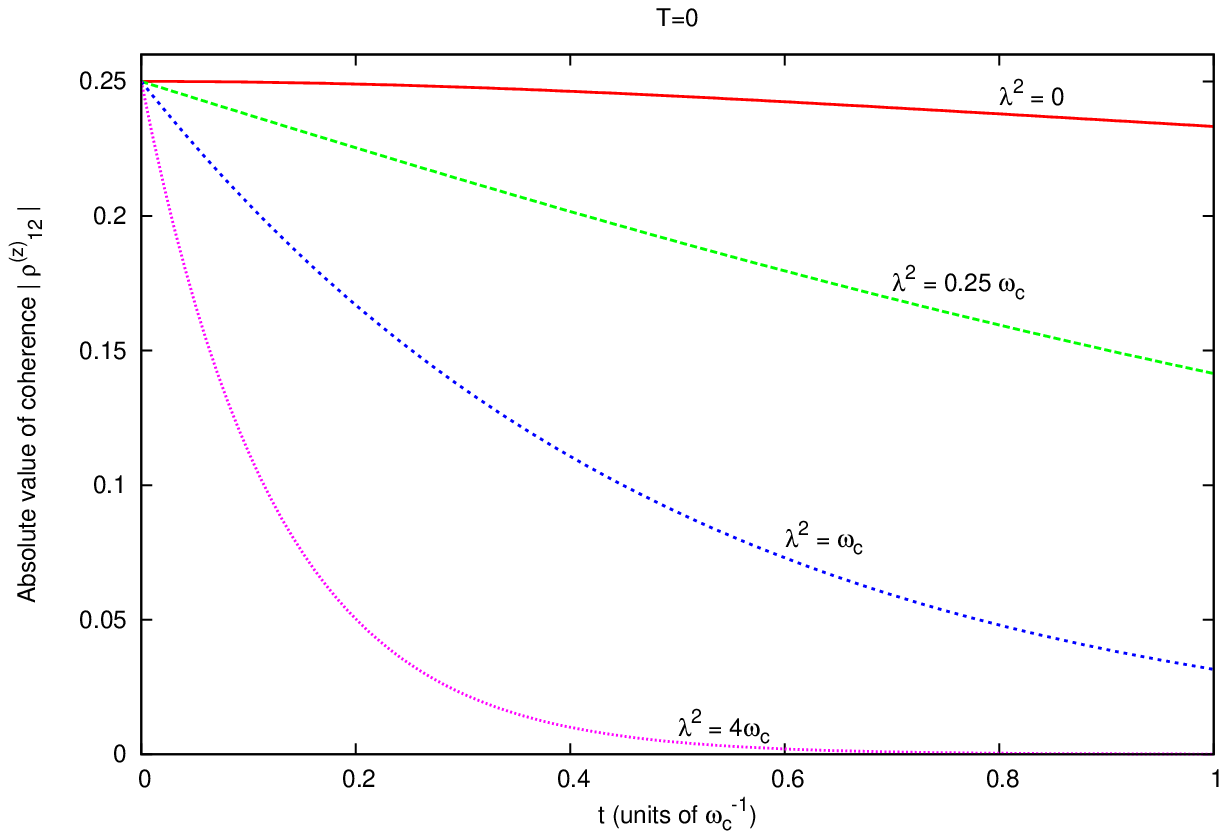}

\includegraphics{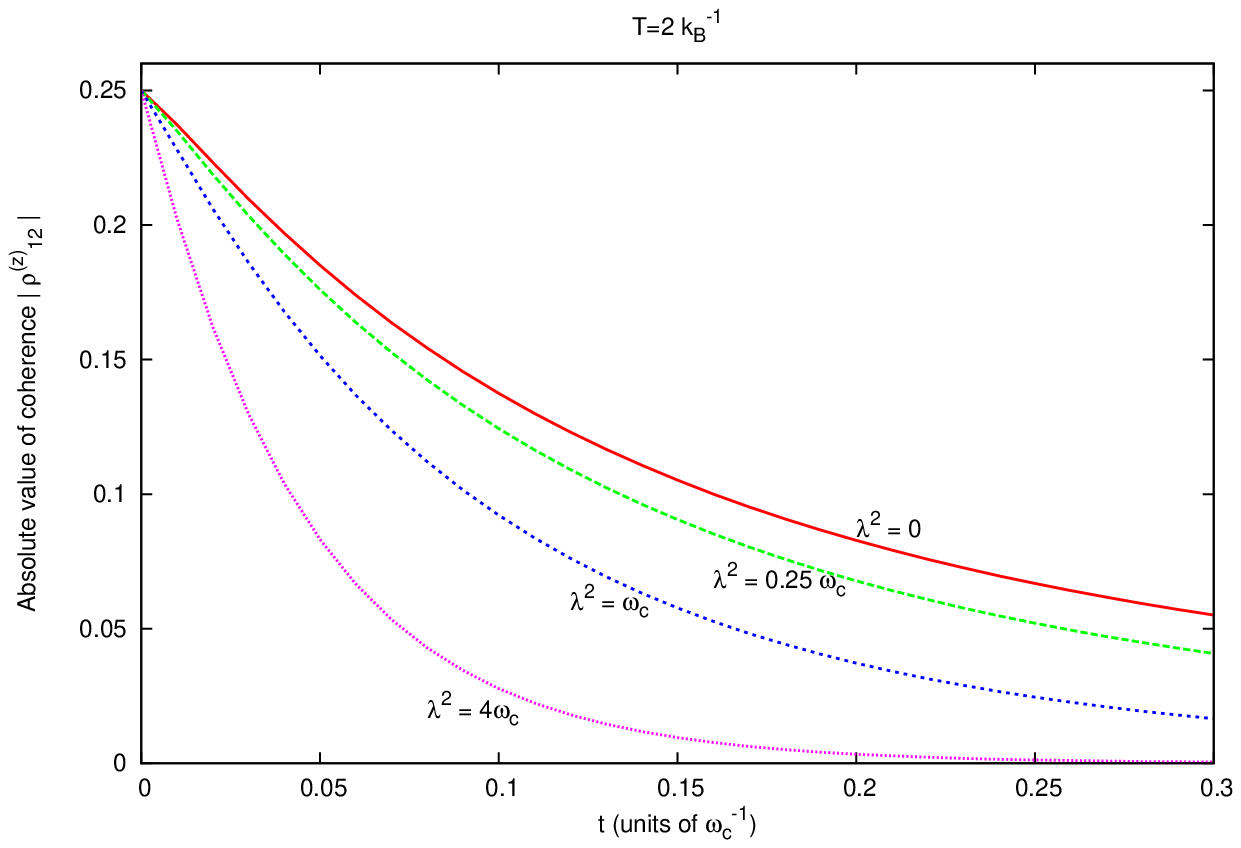}

\caption{(Color online) Time evolution of the absolute value of the coherence
in the eigenbasis of $\hat{\sigma}_{z}$ of a system suffering phase
noise from an Ohmic environment ($\eta=0.05$), while the observable
$\hat{\sigma}_{z}$ is continuously measured. The curves were calculated
using the analytical result of Eq. (\ref{r12zfinal}). The more the
strength of measurement increases, the faster the coherence goes to
zero. The initial conditions are the same applied in Fig. 1.}
\end{figure}

In short, when the observable measured is $\hat{\sigma}_{z}$, the
measurement apparatus reinforces the effect of the environment, thus
leading to faster decoherence. When the observable is $\hat{\sigma}_{x}$,
the two effects compete against each other, leading to a smaller error
than that expected for a system that is not subject to measurement.

\section{Conclusions and Perspectives}

In this work, we have analyzed a two-state system subject to finite-time
measurements that commute or anti-commute with the interaction Hamiltonian
which couples the system with a phase-noisy environment. In the first
case, we have shown that the complete analytical results for a finite-time
measurement in any temperature - seen in Eqs. (\ref{popzfinal}),
(\ref{ro12z}) and (\ref{r12zfinal}) - give only faster rates of
decoherence as the strength of the measurement grows. In the latter
case, we have arrived at an analytical expression for zero temperature
in Eq. (\ref{roxfinal}) which, together with the numerical method
in Eq. (\ref{eq:algorithm3}), has made it possible to conclude that
the initial state of the system under consideration is protected by
the measurement. This result, a consequence of the fact that two simultaneous
measurements of non-commuting observables interfere with each other,
allows a deeper understanding of the measurement process by opening
some interesting perspectives in the area of protection against errors
and demonstrating that it is necessary to take into account that the
measurement is a continuous process rather than an instantaneous collapse
in order to have reliable estimates of the error.

Our approach of considering the system interacting with the environment,
but the apparatus interacting only with the system, through our treatment
of the Lindblad equation, is novel.\textcolor{green}{{} }However, we
have treated the problem with a series of simplifications, namely:
zero temperature for the initial state of the environment, Ohmic density
of states, and the neglect of the system Hamiltonian when measuring
an observable that does not commute with it. Subsequent studies may
use our method in more general cases, even varying the number of states
of the system considered. A first step in the direction of studying
systems interacting with environments initially at temperature $T>0$
was given in Eq. (\ref{Tnnulo}), which shows the differential equations
for the case of finite temperature that may serve for numerical resolutions
that could reveal new effects.

The present formalism does not allow us to approach the possibility
of the state measured being different from an eigenvalue of the system,
as considered in some works on quantum theory of measurement \cite{key-40}.
However, it is still open the possibility of expanding our method
for non-Markovian systems, time-dependent Nakajima-Zwanzig operators,
and to analyze the quantum Zeno effect \cite{key-13,key-14,key-15}.
In particular, the quantum Zeno effect, viewed as a dynamical phenomenon
\cite{key-47}, is closely connected with the fact that a finite-time
measurement protects the initial state in the case of a measured observable
that does not commute with the perturbing Hamiltonian. Roughly speaking,
a finite-time measurement could be thought of as an approximation
for a sequence of repeated instantaneous measurements, as in the bang-bang
decoupling method \cite{key-48}, which has been generalized to arbitrarily
fast and strong pulse sequences, requiring no symmetry. Therefore,
we expect to be able to connect the findings of Ref. \cite{key-48}
with our present approach, clarifying how a general, finite-time measurement
dynamically induces the quantum Zeno effect to protect the initial
state being measured.
\begin{acknowledgments}
C. A. Brasil acknowledges support from Coordenação de Aperfeiçoamento
de Pessoal de Nível Superior (CAPES), Brazil.

L. A. de Castro acknowledges support from Fundação de Amparo à Pesquisa
do Estado de São Paulo (FAPESP), Brazil, project number 2009/12460-0.

R. d. J. Napolitano acknowledges support from Conselho Nacional de
Desenvolvimento Científi{}co e Tecnológico (CNPq), Brazil.
\end{acknowledgments}

\section*{Appendix A: Calculation of the environment degrees of freedom (Sec.
iii-b)}

Using Eq. (\ref{solsupexpB}), we can write:

\begin{eqnarray}
\left(I\right) & \mathrm{Tr}_{B}\left\{ e^{-\hat{\hat{B}}t}\hat{B}\left\{ e^{\hat{\hat{B}}\left(t-t'\right)}\left[\left(e^{\hat{\hat{B}}t'}\hat{\rho}_{B}\right)\hat{B}\right]\right\} \right\}  & =\mathrm{Tr}_{B}\left\{ e^{i\frac{\hat{H}_{B}}{\hbar}t}\hat{B}e^{-i\frac{\hat{H}_{B}}{\hbar}t}\hat{\rho}_{B}e^{i\frac{\hat{H}_{B}}{\hbar}t'}\hat{B}e^{-i\frac{\hat{H}_{B}}{\hbar}t'}\right\} ,\label{termoI}
\end{eqnarray}
\[
\]

\begin{eqnarray}
\left(II\right) & \mathrm{Tr}_{B}\left\{ e^{-\hat{\hat{B}}t}\left\{ e^{\hat{\hat{B}}\left(t-t'\right)}\left[\left(e^{\hat{\hat{B}}t'}\hat{\rho}_{B}\right)\hat{B}\right]\right\} \hat{B}\right\} = & \mathrm{Tr}_{B}\left\{ e^{i\frac{\hat{H}_{B}}{\hbar}t}\hat{B}e^{-i\frac{\hat{H}_{B}}{\hbar}t}\hat{\rho}_{B}e^{i\frac{\hat{H}_{B}}{\hbar}t'}\hat{B}e^{-i\frac{\hat{H}_{B}}{\hbar}t'}\right\} ,\label{termoII}
\end{eqnarray}
\[
\]
\begin{eqnarray}
\left(III\right) & \mathrm{Tr}_{B}\left\{ e^{-\hat{\hat{B}}t}\hat{B}\left\{ e^{\hat{\hat{B}}\left(t-t'\right)}\left[\hat{B}\left(e^{\hat{\hat{B}}t'}\hat{\rho}_{B}\right)\right]\right\} \right\} = & \mathrm{Tr}_{B}\left\{ e^{i\frac{\hat{H}_{B}}{\hbar}t'}\hat{B}e^{-i\frac{\hat{H}_{B}}{\hbar}t'}\hat{\rho}_{B}e^{i\frac{\hat{H}_{B}}{\hbar}t}\hat{B}e^{-i\frac{\hat{H}_{B}}{\hbar}t}\right\} ,\label{termoIII}
\end{eqnarray}

\[
\]
\begin{eqnarray}
\left(IV\right) & \mathrm{Tr}_{B}\left\{ e^{-\hat{\hat{B}}t}\left\{ e^{\hat{\hat{B}}\left(t-t'\right)}\left[\hat{B}\left(e^{\hat{\hat{B}}t'}\hat{\rho}_{B}\right)\right]\right\} \hat{B}\right\} = & \mathrm{Tr}_{B}\left\{ e^{i\frac{\hat{H}_{B}}{\hbar}t'}\hat{B}e^{-i\frac{\hat{H}_{B}}{\hbar}t'}\hat{\rho}_{B}e^{i\frac{\hat{H}_{B}}{\hbar}t}\hat{B}e^{-i\frac{\hat{H}_{B}}{\hbar}t}\right\} .\label{termoIV}
\end{eqnarray}
Let us insert, then, Eqs. (\ref{termoI}), (\ref{termoII}), (\ref{termoIII}),
and (\ref{termoIV}) into Eq. (\ref{intsep}) and group the similar
terms:

\begin{eqnarray}
\hat{\hat{P}}\hat{\hat{G}}\left(t\right)\hat{\hat{G}}\left(t'\right)\hat{\hat{P}}\hat{\alpha}\left(t\right) & = & \left\{ e^{-\hat{\hat{S}}t}\hat{\sigma}_{z}\left\{ e^{\hat{\hat{S}}\left(t-t'\right)}\left[\left(e^{\hat{\hat{S}}t'}\hat{R}\left(t\right)\right)\hat{\sigma}_{z}\right]\right\} \right.\nonumber \\
 & - & \left.e^{-\hat{\hat{S}}t}\left\{ e^{\hat{\hat{S}}\left(t-t'\right)}\left[\left(e^{\hat{\hat{S}}t'}\hat{R}\left(t\right)\right)\hat{\sigma}_{z}\right]\right\} \hat{\sigma}_{z}\right\} \nonumber \\
 & \times & \mathrm{Tr}_{B}\left\{ e^{i\frac{\hat{H}_{B}}{\hbar}t}\hat{B}e^{-i\frac{\hat{H}_{B}}{\hbar}t}\hat{\rho}_{B}e^{i\frac{\hat{H}_{B}}{\hbar}t'}\hat{B}e^{-i\frac{\hat{H}_{B}}{\hbar}t'}\right\} \otimes\hat{\rho}_{B}\nonumber \\
\nonumber \\
 & + & \left\{ e^{-\hat{\hat{S}}t}\left\{ e^{\hat{\hat{S}}\left(t-t'\right)}\left[\hat{\sigma}_{z}\left(e^{\hat{\hat{S}}t'}\hat{R}\left(t\right)\right)\right]\right\} \hat{\sigma}_{z}\right.\nonumber \\
 & - & \left.e^{-\hat{\hat{S}}t}\hat{\sigma}_{z}\left\{ e^{\hat{\hat{S}}\left(t-t'\right)}\left[\hat{\sigma}_{z}\left(e^{\hat{\hat{S}}t'}\hat{R}\left(t\right)\right)\right]\right\} \right\} \nonumber \\
 & \times & \mathrm{Tr}_{B}\left\{ e^{i\frac{\hat{H}_{B}}{\hbar}t'}\hat{B}e^{-i\frac{\hat{H}_{B}}{\hbar}t'}\hat{\rho}_{B}e^{i\frac{\hat{H}_{B}}{\hbar}t}\hat{B}e^{-i\frac{\hat{H}_{B}}{\hbar}t}\right\} \otimes\hat{\rho}_{B}.\label{inttrac}
\end{eqnarray}

The two terms in Eq. (\ref{inttrac}) involving the trace over the
environmental degrees of freedom have the same form. Then, it is sufficient
to evaluate the first of them and perform the change
\[
\begin{cases}
t & \rightarrow t'\\
t' & \rightarrow t
\end{cases}
\]
to obtain the second. To take the partial trace, we use the \emph{Fock-state
basis} $\left|n_{1}\right\rangle \otimes\left|n_{2}\right\rangle ...\equiv\underset{m}{\prod}\left|n_{m}\right\rangle $.
Thus,

\[
\mathrm{Tr}_{B}\left\{ \right\} \rightarrow\underset{n_{1},n_{2},...}{\sum}\left[\underset{m}{\prod}\left\langle n_{m}\right|\right]\left[\underset{m}{\prod}\left|n_{m}\right\rangle \right]
\]
and we obtain \cite{key-42}

\begin{equation}
e^{it\underset{l}{\sum}\omega_{l}\hat{b}_{l}^{\dagger}\hat{b}_{l}}\underset{k}{\sum}\left(g_{k}^{*}\hat{b}_{k}+g_{k}\hat{b}_{k}^{\dagger}\right)e^{-it\underset{l}{\sum}\omega_{l}\hat{b}_{l}^{\dagger}\hat{b}_{l}}=\underset{k}{\sum}\left(g_{k}^{*}\hat{b}_{k}e^{-i\omega_{k}t}+g_{k}\hat{b}_{k}^{\dagger}e^{i\omega_{k}t}\right).\label{aux17}
\end{equation}
if we specify $\hat{H}_{B}$ and $\hat{B}$ as given by Eqs. (\ref{Hb})
and (\ref{opbanho}). By Eq. (\ref{aux17}), the trace over the environmental
degrees of freedom becomes

\begin{eqnarray*}
\mathrm{Tr}_{B}\left\{ e^{i\frac{\hat{H}_{B}}{\hbar}t}\hat{B}e^{-i\frac{\hat{H}_{B}}{\hbar}t}\hat{\rho}_{B}e^{i\frac{\hat{H}_{B}}{\hbar}t'}\hat{B}e^{-i\frac{\hat{H}_{B}}{\hbar}t'}\right\}  & = & \underset{n_{1},n_{2},...}{\sum}\left[\underset{m}{\prod}\left\langle n_{m}\right|\right]\underset{l}{\sum}\left(g_{l}^{*}\hat{b}_{l}e^{-i\omega_{l}t}+g_{l}\hat{b}_{l}^{\dagger}e^{i\omega_{l}t}\right)\\
 & \times & \hat{\rho}_{B}\underset{k}{\sum}\left(g_{k}^{*}\hat{b}_{k}e^{-i\omega_{k}t'}+g_{k}\hat{b}_{k}^{\dagger}e^{i\omega_{k}t'}\right)\left[\underset{m}{\prod}\left|n_{m}\right\rangle \right].
\end{eqnarray*}
With (\ref{rBtermico}), we will obtain

\begin{eqnarray*}
\mathrm{Tr}_{B}\left\{ e^{i\frac{\hat{H}_{B}}{\hbar}t}\hat{B}e^{-i\frac{\hat{H}_{B}}{\hbar}t}\hat{\rho}_{B}e^{i\frac{\hat{H}_{B}}{\hbar}t'}\hat{B}e^{-i\frac{\hat{H}_{B}}{\hbar}t'}\right\}  & = & \underset{l}{\sum}\left|g_{l}\right|^{2}\left[e^{i\omega_{l}\left(t-t'\right)}+\frac{e^{i\omega_{l}\left(t-t'\right)}}{e^{\hbar\beta\omega_{l}}-1}+\frac{e^{-i\omega_{l}\left(t-t'\right)}}{e^{\hbar\beta\omega_{l}}-1}\right]\\
 & = & \underset{l}{\sum}\left|g_{l}\right|^{2}\left[e^{i\omega_{l}\left(t-t'\right)}+\frac{2\cos\left[\omega_{l}\left(t-t'\right)\right]}{e^{\hbar\beta\omega_{l}}-1}\right],
\end{eqnarray*}
or, more simply,

\begin{eqnarray}
\mathrm{Tr}_{B}\left\{ e^{i\frac{\hat{H}_{B}}{\hbar}t}\hat{B}e^{-i\frac{\hat{H}_{B}}{\hbar}t}\hat{\rho}_{B}e^{i\frac{\hat{H}_{B}}{\hbar}t'}\hat{B}e^{-i\frac{\hat{H}_{B}}{\hbar}t'}\right\}  & = & \underset{l}{\sum}\left|g_{l}\right|^{2}\left\{ \coth\left(\frac{\hbar\beta\omega_{l}}{2}\right)\cos\left[\omega_{l}\left(t-t'\right)\right]\right.\nonumber \\
 & + & \left.i\sin\left[\omega_{l}\left(t-t'\right)\right]\right\} .\label{tramb}
\end{eqnarray}

As explained above, the change of varibles
\[
\begin{cases}
t & \rightarrow t'\\
t' & \rightarrow t
\end{cases}
\]
 in Eq. (\ref{tramb}) gives us the other necessary term for Eq. (\ref{inttrac}),
that is,

\begin{eqnarray}
\mathrm{Tr}_{B}\left\{ e^{i\frac{\hat{H}_{B}}{\hbar}t'}\hat{B}e^{-i\frac{\hat{H}_{B}}{\hbar}t'}\hat{\rho}_{B}e^{i\frac{\hat{H}_{B}}{\hbar}t}\hat{B}e^{-i\frac{\hat{H}_{B}}{\hbar}t}\right\}  & = & \underset{l}{\sum}\left|g_{l}\right|^{2}\left\{ \coth\left(\frac{\hbar\beta\omega_{l}}{2}\right)\cos\left[\omega_{l}\left(t-t'\right)\right]\right.\nonumber \\
 & - & \left.i\sin\left[\omega_{l}\left(t-t'\right)\right]\right\} .\label{aux22}
\end{eqnarray}
The final result, Eq. (\ref{ambfinal}), is achieved by inserting
Eqs. (\ref{tramb}) and (\ref{aux22}) into Eq. (\ref{inttrac}).

\section*{Appendix B: the case of $\hat{L}^{\left(S\right)}=\lambda\hat{\sigma}_{z}$
(Sec. III-E)}

Using Eq. (\ref{R(t)}) we can evaluate the terms $\left(A\right)$,
$\left(B\right)$, $\left(C\right)$, and $\left(D\right)$ of Eq.
(\ref{ambcont}). Therefore,

\begin{eqnarray}
\left(A\right) & e^{-\hat{\hat{S}}t}\hat{\sigma}_{z}\left\{ e^{\hat{\hat{S}}\left(t-t'\right)}\left[\left(e^{\hat{\hat{S}}t'}\hat{R}\left(t\right)\right)\hat{\sigma}_{z}\right]\right\} = & \left(\begin{array}{cc}
R_{11} & -R_{12}\\
-R_{21} & R_{22}
\end{array}\right),\label{termoA}
\end{eqnarray}

\[
\]

\begin{eqnarray}
\left(B\right) & e^{-\hat{\hat{S}}t}\left\{ e^{\hat{\hat{S}}\left(t-t'\right)}\left[\left(e^{\hat{\hat{S}}t'}\hat{R}\left(t\right)\right)\hat{\sigma}_{z}\right]\right\} \hat{\sigma}_{z} & =\left(\begin{array}{cc}
R_{11} & R_{12}\\
R_{21} & R_{22}
\end{array}\right),\label{termoB}
\end{eqnarray}
\[
\]

\begin{eqnarray}
\left(C\right) & e^{-\hat{\hat{S}}t}\left\{ e^{\hat{\hat{S}}\left(t-t'\right)}\left[\hat{\sigma}_{z}\left(e^{\hat{\hat{S}}t'}\hat{R}\left(t\right)\right)\right]\right\} \hat{\sigma}_{z} & =\left(\begin{array}{cc}
R_{11} & -R_{12}\\
-R_{21} & R_{22}
\end{array}\right),\label{termoC}
\end{eqnarray}
and
\[
\]

\begin{eqnarray}
\left(D\right) & e^{-\hat{\hat{S}}t}\hat{\sigma}_{z}\left\{ e^{\hat{\hat{S}}\left(t-t'\right)}\left[\hat{\sigma}_{z}\left(e^{\hat{\hat{S}}t'}\hat{R}\left(t\right)\right)\right]\right\}  & =\left(\begin{array}{cc}
R_{11} & R_{12}\\
R_{21} & R_{22}
\end{array}\right).\label{termoD}
\end{eqnarray}

With Eqs. (\ref{termoA}), (\ref{termoB}), (\ref{termoC}), and (\ref{termoD}),
we can simplify Eq. (\ref{ambcont}):
\begin{eqnarray}
\hat{\hat{P}}\hat{\hat{G}}\left(t\right)\hat{\hat{G}}\left(t'\right)\hat{\hat{P}}\hat{\alpha}\left(t\right) & = & -4\int_{0}^{\infty}d\omega J\left(\omega\right)\left(\begin{array}{cc}
0 & R_{12}\\
R_{21} & 0
\end{array}\right)\cos\left[\omega\left(t-t'\right)\right]\coth\left(\frac{\beta\hbar\omega}{2}\right)\otimes\hat{\rho}_{B}.\label{intz}
\end{eqnarray}
Hence, we substitute Eq. (\ref{intz}) into the original Eq. (\ref{eqFINAL})
and obtain Eq. (\ref{eqzcont}).

\section*{Appendix C: the coherence for the case of $\hat{L}^{\left(S\right)}=\lambda\hat{\sigma}_{z}$
(Sec. III-E.2)}

In (\ref{eqdifcoerz}), for the sake of convenience, we perform the
change of variable
\begin{eqnarray*}
\tau & = & t-t',\\
d\tau & = & -dt',
\end{eqnarray*}
with

\begin{eqnarray*}
\int_{0}^{t}dt' & = & -\int_{t}^{0}d\tau=\int_{0}^{t}d\tau,
\end{eqnarray*}
that gives
\[
\frac{d}{dt}R_{ij}=-4\eta R_{ij}\left(t\right)\int_{0}^{t}d\tau\int_{0}^{\infty}d\omega\omega e^{-\frac{\omega}{\omega_{c}}}\cos\left(\omega\tau\right)\coth\left(\frac{\beta\hbar\omega}{2}\right).
\]
The first step is to solve the time integral, that leads us to
\[
\frac{d}{dt}R_{ij}=-4\eta R_{ij}\left(t\right)\int_{0}^{\infty}d\omega e^{-\frac{\omega}{\omega_{c}}}\sin\left(\omega t\right)\coth\left(\frac{\beta\hbar\omega}{2}\right).
\]
The frequency integral, evaluated with the help of Ref. \cite{key-43},
gives

\begin{eqnarray*}
\frac{d}{dt}R_{ij} & = & i2\frac{\eta}{\beta\hbar}R_{ij}\left(t\right)\left[\psi\left(\frac{1}{\omega_{c}\beta\hbar}+i\frac{t}{\beta\hbar}\right)-\psi\left(\frac{1}{\omega_{c}\beta\hbar}-i\frac{t}{\beta\hbar}\right)\right.\\
 & + & \left.\psi\left(\frac{1}{\omega_{c}\beta\hbar}+1+i\frac{t}{\beta\hbar}\right)-\psi\left(\frac{1}{\omega_{c}\beta\hbar}+1-i\frac{t}{\beta\hbar}\right)\right],
\end{eqnarray*}
where

\begin{eqnarray*}
\psi\left(z\right) & = & \frac{d}{dz}\ln\left[\Gamma\left(z\right)\right]\:,\mathrm{Re}\left\{ z\right\} >0.
\end{eqnarray*}
The solution is, therefore,
\begin{equation}
\begin{cases}
R_{12}\left(t\right)= & R_{12}\left(0\right)\left[\frac{\Gamma\left(\frac{1}{\omega_{c}\beta\hbar}+i\frac{t}{\beta\hbar}\right)\Gamma\left(\frac{1}{\omega_{c}\beta\hbar}-i\frac{t}{\beta\hbar}\right)}{\Gamma^{2}\left(\frac{1}{\omega_{c}\beta\hbar}\right)}\frac{\Gamma\left(\frac{1}{\omega_{c}\beta\hbar}+1+i\frac{t}{\beta\hbar}\right)\Gamma\left(\frac{1}{\omega_{c}\beta\hbar}+1-i\frac{t}{\beta\hbar}\right)}{\Gamma^{2}\left(\frac{1}{\omega_{c}\beta\hbar}+1\right)}\right]^{2\eta},\\
R_{21}\left(t\right)= & R_{21}\left(0\right)\left[\frac{\Gamma\left(\frac{1}{\omega_{c}\beta\hbar}+i\frac{t}{\beta\hbar}\right)\Gamma\left(\frac{1}{\omega_{c}\beta\hbar}-i\frac{t}{\beta\hbar}\right)}{\Gamma^{2}\left(\frac{1}{\omega_{c}\beta\hbar}\right)}\frac{\Gamma\left(\frac{1}{\omega_{c}\beta\hbar}+1+i\frac{t}{\beta\hbar}\right)\Gamma\left(\frac{1}{\omega_{c}\beta\hbar}+1-i\frac{t}{\beta\hbar}\right)}{\Gamma^{2}\left(\frac{1}{\omega_{c}\beta\hbar}+1\right)}\right]^{2\eta}.
\end{cases}\label{Rndzfinal}
\end{equation}
The coherences are obtained by performing the transformation of Eq.
(\ref{roinv}) to Eq. (\ref{Rndzfinal}). Since $\rho_{21}\left(t\right)=\rho_{12}^{*}\left(t\right)$,
we consider only the element $\rho_{12}\left(t\right)$ and the result
is Eq. (\ref{ro12z}).

\section*{Appendix D: Simplifying the expression of the coherence for the case
of $\hat{L}^{\left(S\right)}=\lambda\hat{\sigma}_{z}$ (Sec. III-E.2)}

Equation (\ref{ro12z}) can be simplified by using the properties
of the gamma function \cite{key-43}. With $\Gamma\left(x+1\right)=x\Gamma\left(x\right)$,
we obtain

\[
\rho_{12}\left(t\right)=\rho_{12}\left(0\right)\left\{ \frac{\Gamma^{2}\left(\frac{1}{\omega_{c}\beta\hbar}+i\frac{t}{\beta\hbar}\right)\Gamma^{2}\left(\frac{1}{\omega_{c}\beta\hbar}-i\frac{t}{\beta\hbar}\right)}{\Gamma^{4}\left(\frac{1}{\omega_{c}\beta\hbar}\right)}\left[1+\left(\omega_{c}t\right)^{2}\right]\right\} ^{2\eta}e^{-2\lambda^{2}t}e^{i2\omega_{0}t}.
\]
Now, using the product representation
\begin{eqnarray*}
 & \Gamma\left(z\right)=e^{-Cz}\frac{1}{z}\underset{k=1}{\overset{\infty}{\prod}}\frac{e^{\frac{z}{k}}}{1+\frac{z}{k}},\:\mathrm{Re}\left\{ z\right\} >0,
\end{eqnarray*}
where $C$ is the \emph{Euler constant,}

\begin{eqnarray*}
C= & \underset{s\rightarrow\infty}{\lim}\left[\underset{m=1}{\overset{s}{\sum}}\frac{1}{m}-ln\left(s\right)\right] & =0.577215...,
\end{eqnarray*}
we see that
\begin{eqnarray*}
 & \frac{\Gamma\left(\frac{1}{\omega_{c}\beta\hbar}+i\frac{t}{\beta\hbar}\right)\Gamma\left(\frac{1}{\omega_{c}\beta\hbar}-i\frac{t}{\beta\hbar}\right)}{\Gamma^{2}\left(\frac{1}{\omega_{c}\beta\hbar}\right)}=\frac{1}{1+\left(\omega_{c}t\right)^{2}}\underset{n=1}{\overset{\infty}{\prod}}\frac{n\omega_{c}\beta\hbar+1}{1+n\omega_{c}\beta\hbar+\omega_{c}t},
\end{eqnarray*}
resulting, finally, in Eq. (\ref{r12zfinal}), which is suitable for
numerical implementations.

\section*{Appendix E: the differential equations for the case of $\hat{L}^{\left(S\right)}=\lambda\hat{\sigma}_{x}$
(Sec. III-F)}

Using the definitions (\ref{aux29}) and (\ref{aux30}) over (\ref{Rx}),
we have

\begin{eqnarray*}
e^{\hat{\hat{S}}t'}\hat{R}\left(t\right) & = & e^{\hat{\hat{S}}t'}\left(\begin{array}{cc}
r_{11} & r_{12}\\
r_{21} & r_{22}
\end{array}\right)=\left(\begin{array}{cc}
r_{11}^{\left(0\right)} & r_{12}^{\left(0\right)}\\
r_{21}^{\left(0\right)} & r_{22}^{\left(0\right)}
\end{array}\right),
\end{eqnarray*}
where

\begin{equation}
\begin{cases}
r_{11}^{\left(0\right)}= & \frac{r_{11}-r_{22}}{2}e^{-2\lambda^{2}t'}+\frac{r_{11}+r_{22}}{2},\\
r_{22}^{\left(0\right)}= & -\frac{r_{11}-r_{22}}{2}e^{-2\lambda^{2}t'}+\frac{r_{11}+r_{22}}{2},
\end{cases}\label{r0pop}
\end{equation}
and

\begin{equation}
\begin{cases}
r_{12}^{\left(0\right)}= & \frac{e^{-\lambda^{2}t'}}{\Omega}\left[\Omega\cosh\left(\Omega t'\right)r_{12}-i2\omega_{0}\sinh\left(\Omega t'\right)r_{12}+\lambda^{2}\sinh\left(\Omega t'\right)r_{21}\right],\\
r_{21}^{\left(0\right)}= & \frac{e^{-\lambda^{2}t'}}{\Omega}\left[\Omega\cosh\left(\Omega t'\right)r_{21}+i2\omega_{0}\sinh\left(\Omega t'\right)r_{21}+\lambda^{2}\sinh\left(\Omega t'\right)r_{12}\right].
\end{cases}\label{r0coer}
\end{equation}
Now we can evaluate the terms between brackets on (\ref{ambcont}),
that will furnish

\begin{eqnarray*}
e^{\hat{\hat{S}}\left(t-t'\right)}\left[\hat{\sigma}_{z}\left(e^{\hat{\hat{S}}t'}\hat{R}\left(t\right)\right)\right] & = & \left(\begin{array}{cc}
r_{11}^{\left(1\right)} & r_{12}^{\left(1\right)}\\
r_{21}^{\left(1\right)} & r_{22}^{\left(1\right)}
\end{array}\right)
\end{eqnarray*}
and
\begin{eqnarray*}
e^{\hat{\hat{S}}\left(t-t'\right)}\left[\left(e^{\hat{\hat{S}}t'}\hat{R}\left(t\right)\right)\hat{\sigma}_{z}\right] & = & \left(\begin{array}{cc}
r_{11}^{\left(2\right)} & r_{12}^{\left(2\right)}\\
r_{21}^{\left(2\right)} & r_{22}^{\left(2\right)}
\end{array}\right)
\end{eqnarray*}
where, using (\ref{r0pop}) and (\ref{r0coer}),

\begin{eqnarray*}
 & \begin{cases}
r_{11}^{\left(1\right)}= & \frac{r_{11}^{\left(0\right)}+r_{22}^{\left(0\right)}}{2}e^{-2\lambda^{2}\left(t-t'\right)}+\frac{r_{11}^{\left(0\right)}-r_{22}^{\left(0\right)}}{2},\\
r_{22}^{\left(1\right)}= & -\frac{r_{11}^{\left(0\right)}+r_{22}^{\left(0\right)}}{2}e^{-2\lambda^{2}\left(t-t'\right)}+\frac{r_{11}^{\left(0\right)}-r_{22}^{\left(0\right)}}{2},
\end{cases}
\end{eqnarray*}

\begin{eqnarray*}
 & \begin{cases}
r_{12}^{\left(1\right)}= & \frac{e^{-\lambda^{2}\left(t-t'\right)}}{\Omega}\left\{ \Omega\cosh\left[\Omega\left(t-t'\right)\right]r_{12}^{\left(0\right)}-i2\omega_{0}\sinh\left[\Omega\left(t-t'\right)\right]r_{12}^{\left(0\right)}-\lambda^{2}\sinh\left[\Omega\left(t-t'\right)\right]r_{21}^{\left(0\right)}\right\} ,\\
r_{21}^{\left(1\right)}= & -\frac{e^{-\lambda^{2}\left(t-t'\right)}}{\Omega}\left\{ \Omega\cosh\left[\Omega\left(t-t'\right)\right]r_{21}^{\left(0\right)}+i2\omega_{0}\sinh\left[\Omega\left(t-t'\right)\right]r_{21}^{\left(0\right)}-\lambda^{2}\sinh\left[\Omega\left(t-t'\right)\right]r_{12}^{\left(0\right)}\right\} ,
\end{cases}
\end{eqnarray*}

\begin{eqnarray*}
 & \begin{cases}
r_{11}^{\left(2\right)}= & \frac{r_{11}^{\left(0\right)}+r_{22}^{\left(0\right)}}{2}e^{-2\lambda^{2}\left(t-t'\right)}+\frac{r_{11}^{\left(0\right)}-r_{22}^{\left(0\right)}}{2},\\
r_{22}^{\left(2\right)}= & -\frac{r_{11}^{\left(0\right)}+r_{22}^{\left(0\right)}}{2}e^{-2\lambda^{2}\left(t-t'\right)}+\frac{r_{11}^{\left(0\right)}-r_{22}^{\left(0\right)}}{2},
\end{cases}
\end{eqnarray*}
and

\begin{eqnarray*}
 & \begin{cases}
r_{12}^{\left(2\right)}= & -\frac{e^{-\lambda^{2}\left(t-t'\right)}}{\Omega}\left\{ \Omega\cosh\left[\Omega\left(t-t'\right)\right]r_{12}^{\left(0\right)}-i2\omega_{0}\sinh\left[\Omega\left(t-t'\right)\right]r_{12}^{\left(0\right)}-\lambda^{2}\sinh\left[\Omega\left(t-t'\right)\right]r_{21}^{\left(0\right)}\right\} ,\\
r_{21}^{\left(2\right)}= & \frac{e^{-\lambda^{2}\left(t-t'\right)}}{\Omega}\left\{ \Omega\cosh\left[\Omega\left(t-t'\right)\right]r_{21}^{\left(0\right)}+i2\omega_{0}\sinh\left[\Omega\left(t-t'\right)\right]r_{21}^{\left(0\right)}-\lambda^{2}\sinh\left[\Omega\left(t-t'\right)\right]r_{12}^{\left(0\right)}\right\} ,
\end{cases}
\end{eqnarray*}
i.e., $\begin{cases}
r_{11}^{\left(1\right)} & =r_{11}^{\left(2\right)}\\
r_{22}^{\left(1\right)} & =r_{22}^{\left(2\right)}
\end{cases}$and $\begin{cases}
r_{12}^{\left(1\right)} & =-r_{12}^{\left(2\right)}\\
r_{21}^{\left(1\right)} & =-r_{21}^{\left(2\right)}
\end{cases}$.

The progressive calculus of $\left(A\right)$, $\left(B\right)$,
$\left(C\right)$ and $\left(D\right)$, then, will furnish:
\begin{eqnarray}
\left(A\right) &  & e^{-\hat{\hat{S}}t}\hat{\sigma}_{z}\left\{ e^{\hat{\hat{S}}\left(t-t'\right)}\left[\left(e^{\hat{\hat{S}}t'}\hat{R}\left(t\right)\right)\hat{\sigma}_{z}\right]\right\} =\left(\begin{array}{cc}
r_{11}^{\left(3\right)} & r_{12}^{\left(3\right)}\\
r_{21}^{\left(3\right)} & r_{22}^{\left(3\right)}
\end{array}\right)\label{termoAx}
\end{eqnarray}
with
\begin{eqnarray*}
 &  & \begin{cases}
r_{11}^{\left(3\right)}= & \frac{r_{11}^{\left(2\right)}+r_{22}^{\left(2\right)}}{2}e^{2\lambda^{2}t}+\frac{r_{11}^{\left(2\right)}-r_{22}^{\left(2\right)}}{2},\\
r_{22}^{\left(3\right)}= & -\frac{r_{11}^{\left(2\right)}+r_{22}^{\left(2\right)}}{2}e^{2\lambda^{2}t}+\frac{r_{11}^{\left(2\right)}-r_{22}^{\left(2\right)}}{2},
\end{cases}
\end{eqnarray*}
and

\begin{eqnarray*}
 &  & \begin{cases}
r_{12}^{\left(3\right)}= & \frac{e^{\lambda^{2}t}}{\Omega}\left[\Omega\cosh\left(\Omega t\right)r_{12}^{\left(2\right)}+i2\omega_{0}\sinh\left(\Omega t\right)r_{12}^{\left(2\right)}+\lambda^{2}\sinh\left(\Omega t\right)r_{21}^{\left(2\right)}\right],\\
r_{21}^{\left(3\right)}= & -\frac{e^{\lambda^{2}t}}{\Omega}\left[\Omega\cosh\left(\Omega t\right)r_{21}^{\left(2\right)}-i2\omega_{0}\sinh\left(\Omega t\right)r_{21}^{\left(2\right)}+\lambda^{2}\sinh\left(\Omega t\right)r_{12}^{\left(2\right)}\right].
\end{cases}
\end{eqnarray*}

\begin{eqnarray}
\left(B\right) &  & e^{-\hat{\hat{S}}t}\left\{ e^{\hat{\hat{S}}\left(t-t'\right)}\left[\left(e^{\hat{\hat{S}}t'}\hat{R}\left(t\right)\right)\hat{\sigma}_{z}\right]\right\} \hat{\sigma}_{z}=\left(\begin{array}{cc}
r_{11}^{\left(4\right)} & r_{12}^{\left(4\right)}\\
r_{21}^{\left(4\right)} & r_{22}^{\left(4\right)}
\end{array}\right)\label{termoBx}
\end{eqnarray}
with

\begin{eqnarray*}
 &  & \begin{cases}
r_{11}^{\left(4\right)}= & \frac{r_{11}^{\left(2\right)}+r_{22}^{\left(2\right)}}{2}e^{2\lambda^{2}t}+\frac{r_{11}^{\left(2\right)}-r_{22}^{\left(2\right)}}{2},\\
r_{22}^{\left(4\right)}= & -\frac{r_{11}^{\left(2\right)}+r_{22}^{\left(2\right)}}{2}e^{2\lambda^{2}t}+\frac{r_{11}^{\left(2\right)}-r_{22}^{\left(2\right)}}{2},
\end{cases}
\end{eqnarray*}
and

\begin{eqnarray*}
 &  & \begin{cases}
r_{12}^{\left(4\right)}= & -\frac{e^{\lambda^{2}t}}{\Omega}\left[\Omega\cosh\left(\Omega t\right)r_{12}^{\left(2\right)}+i2\omega_{0}\sinh\left(\Omega t\right)r_{12}^{\left(2\right)}+\lambda^{2}\sinh\left(\Omega t\right)r_{21}^{\left(2\right)}\right],\\
r_{21}^{\left(4\right)}= & \frac{e^{\lambda^{2}t}}{\Omega}\left[\Omega\cosh\left(\Omega t\right)r_{21}^{\left(2\right)}-i2\omega_{0}\sinh\left(\Omega t\right)r_{21}^{\left(2\right)}+\lambda^{2}\sinh\left(\Omega t\right)r_{12}^{\left(2\right)}\right].
\end{cases}
\end{eqnarray*}
We can observe that $\begin{cases}
r_{11}^{\left(3\right)} & =r_{11}^{\left(4\right)}\\
r_{22}^{\left(3\right)} & =r_{22}^{\left(4\right)}
\end{cases}$and $\begin{cases}
r_{12}^{\left(3\right)} & =-r_{12}^{\left(4\right)}\\
r_{21}^{\left(3\right)} & =-r_{21}^{\left(4\right)}
\end{cases}$.
\begin{eqnarray}
\left(C\right) &  & e^{-\hat{\hat{S}}t}\left\{ e^{\hat{\hat{S}}\left(t-t'\right)}\left[\hat{\sigma}_{z}\left(e^{\hat{\hat{S}}t'}\hat{R}\left(t\right)\right)\right]\right\} \hat{\sigma}_{z}=\left(\begin{array}{cc}
r_{11}^{\left(5\right)} & r_{12}^{\left(5\right)}\\
r_{21}^{\left(5\right)} & r_{22}^{\left(5\right)}
\end{array}\right)\label{termoCx}
\end{eqnarray}
with
\begin{eqnarray*}
 & \begin{cases}
r_{11}^{\left(5\right)}= & \frac{r_{11}^{\left(1\right)}+r_{22}^{\left(1\right)}}{2}e^{2\lambda^{2}t}+\frac{r_{11}^{\left(1\right)}-r_{22}^{\left(1\right)}}{2},\\
r_{22}^{\left(5\right)}= & -\frac{r_{11}^{\left(1\right)}+r_{22}^{\left(1\right)}}{2}e^{2\lambda^{2}t}+\frac{r_{11}^{\left(1\right)}-r_{22}^{\left(1\right)}}{2},
\end{cases}
\end{eqnarray*}
and

\begin{eqnarray*}
 & \begin{cases}
r_{12}^{\left(5\right)}= & -\frac{e^{\lambda^{2}t}}{\Omega}\left[\Omega\cosh\left(\Omega t\right)r_{12}^{\left(1\right)}+i2\omega_{0}\sinh\left(\Omega t\right)r_{12}^{\left(1\right)}+\lambda^{2}\sinh\left(\Omega t\right)r_{21}^{\left(1\right)}\right],\\
r_{21}^{\left(5\right)}= & \frac{e^{\lambda^{2}t}}{\Omega}\left[\Omega\cosh\left(\Omega t\right)r_{21}^{\left(1\right)}-i2\omega_{0}\sinh\left(\Omega t\right)r_{21}^{\left(1\right)}+\lambda^{2}\sinh\left(\Omega t\right)r_{12}^{\left(1\right)}\right].
\end{cases}
\end{eqnarray*}

\begin{eqnarray}
\left(D\right) &  & e^{-\hat{\hat{S}}t}\hat{\sigma}_{z}\left\{ e^{\hat{\hat{S}}\left(t-t'\right)}\left[\hat{\sigma}_{z}\left(e^{\hat{\hat{S}}t'}\hat{R}\left(t\right)\right)\right]\right\} =\left(\begin{array}{cc}
r_{11}^{\left(6\right)} & r_{12}^{\left(6\right)}\\
r_{21}^{\left(6\right)} & r_{22}^{\left(6\right)}
\end{array}\right)\label{termoDx}
\end{eqnarray}
with
\begin{eqnarray*}
 & \begin{cases}
r_{11}^{\left(6\right)}= & \frac{r_{11}^{\left(1\right)}+r_{22}^{\left(1\right)}}{2}e^{2\lambda^{2}t}+\frac{r_{11}^{\left(1\right)}-r_{22}^{\left(1\right)}}{2},\\
r_{22}^{\left(6\right)}= & -\frac{r_{11}^{\left(1\right)}+r_{22}^{\left(1\right)}}{2}e^{2\lambda^{2}t}+\frac{r_{11}^{\left(1\right)}-r_{22}^{\left(1\right)}}{2},
\end{cases}
\end{eqnarray*}
and

\begin{eqnarray*}
 & \begin{cases}
r_{12}^{\left(6\right)}= & \frac{e^{\lambda^{2}t}}{\Omega}\left[\Omega\cosh\left(\Omega t\right)r_{12}^{\left(1\right)}+i2\omega_{0}\sinh\left(\Omega t\right)r_{12}^{\left(1\right)}+\lambda^{2}\sinh\left(\Omega t\right)r_{21}^{\left(1\right)}\right],\\
r_{21}^{\left(6\right)}= & -\frac{e^{\lambda^{2}t}}{\Omega}\left[\Omega\cosh\left(\Omega t\right)r_{21}^{\left(1\right)}-i2\omega_{0}\sinh\left(\Omega t\right)r_{21}^{\left(1\right)}+\lambda^{2}\sinh\left(\Omega t\right)r_{12}^{\left(1\right)}\right].
\end{cases}
\end{eqnarray*}
Again, we have another similarity, $\begin{cases}
r_{11}^{\left(5\right)} & =r_{11}^{\left(6\right)}\\
r_{22}^{\left(5\right)} & =r_{22}^{\left(6\right)}
\end{cases}$ and $\begin{cases}
r_{12}^{\left(5\right)} & =-r_{12}^{\left(6\right)}\\
r_{21}^{\left(5\right)} & =-r_{21}^{\left(6\right)}
\end{cases}$. 

By using (\ref{termoAx}), (\ref{termoBx}), (\ref{termoCx}) and
(\ref{termoDx}), with the similarity relations, we have then the
equation (\ref{eqmatfinal}).

\section*{Appendix F: the coherences in the case of $\hat{L}^{\left(S\right)}=\lambda\hat{\sigma}_{x}$
represented in the eigenbasis of $\hat{\sigma}_{z}$, for $T=0$ and
$\omega_{0}=0$ (Sec. III-F.2)}

For $T=0$ and $\omega_{0}=0$, the equation for the coherences becomes
\begin{eqnarray}
\frac{d}{dt}r_{12}\left(t\right) & = & 4\eta\left[\sinh\left(\lambda^{2}t\right)I_{2}\left(t\right)+\cosh\left(\lambda^{2}t\right)I_{3}\left(t\right)\right]r_{21}\left(t\right)\nonumber \\
 & - & 4\eta\left[\cosh\left(\lambda^{2}t\right)I_{2}\left(t\right)+\sinh\left(\lambda^{2}t\right)I_{3}\left(t\right)\right]r_{12}\left(t\right),\label{eqr12aprox}
\end{eqnarray}
with
\begin{eqnarray}
I_{2}\left(t\right) & = & 2\lambda^{2}\sinh\left(\lambda^{2}t\right)g_{0}+\lambda^{2}\sinh\left(\lambda^{2}t\right)g_{1}\left(t\right)+\lambda^{2}\cosh\left(\lambda^{2}t\right)g_{2}\left(t\right),\label{I2aprox}
\end{eqnarray}
\begin{eqnarray}
I_{3}\left(t\right) & = & -2\lambda^{2}\cosh\left(\lambda^{2}t\right)g_{0}+\lambda^{2}\cosh\left(\lambda^{2}t\right)g_{1}\left(t\right)+\lambda^{2}\sinh\left(\lambda^{2}t\right)g_{2}\left(t\right),\label{I3aprox}
\end{eqnarray}
and

\[
g_{0}=\mathrm{Re}\left\{ \exp\left(i\frac{2\lambda^{2}}{\omega_{C}}\right)\Gamma\left(0,i\frac{2\lambda^{2}}{\omega_{C}}\right)\right\} ,
\]

\begin{eqnarray*}
g_{1}\left(t\right) & = & \mathrm{Re}\left\{ \exp\left(2\lambda^{2}t+i\frac{2\lambda^{2}}{\omega_{C}}\right)\Gamma\left(0,2\lambda^{2}t+i\frac{2\lambda^{2}}{\omega_{C}}\right)\right.\\
 & + & \left.\exp\left[-\left(2\lambda^{2}t+i\frac{2\lambda^{2}}{\omega_{C}}\right)\right]\Gamma\left[0,-\left(2\lambda^{2}t+i\frac{2\lambda^{2}}{\omega_{C}}\right)\right]\right\} ,
\end{eqnarray*}
and
\begin{eqnarray*}
g_{2}\left(t\right) & = & \mathrm{Re}\left\{ \exp\left[\left(2\lambda^{2}t+i\frac{2\lambda^{2}}{\omega_{C}}\right)\right]\Gamma\left[-1,\left(2\lambda^{2}t+i\frac{2\lambda^{2}}{\omega_{C}}\right)\right]\right.\\
 & - & \left.\exp\left[-\left(2\lambda^{2}t+i\frac{2\lambda^{2}}{\omega_{C}}\right)\right]\Gamma\left[-1,-\left(2\lambda^{2}t+i\frac{2\lambda^{2}}{\omega_{C}}\right)\right]\right\} .
\end{eqnarray*}
Let us write the complex conjugate of Eq. (\ref{eqr12aprox}):
\begin{eqnarray}
\frac{d}{dt}r_{21}\left(t\right) & = & 4\eta\left[\sinh\left(\lambda^{2}t\right)I_{2}\left(t\right)+\cosh\left(\lambda^{2}t\right)I_{3}\left(t\right)\right]r_{12}\left(t\right)\nonumber \\
 & - & 4\eta\left[\cosh\left(\lambda^{2}t\right)I_{2}\left(t\right)+\sinh\left(\lambda^{2}t\right)I_{3}\left(t\right)\right]r_{21}\left(t\right).\label{eqr21aprox}
\end{eqnarray}
Adding Eqs. (\ref{eqr12aprox}) and (\ref{eqr21aprox}) and subtracting
Eq. (\ref{eqr21aprox}) from (\ref{eqr12aprox}), we obtain the decoupled
system of equations:

\[
\begin{cases}
\frac{d}{dt}\mathrm{Re}\left\{ r_{12}\left(t\right)\right\} = & 4\eta\left[\sinh\left(\lambda^{2}t\right)-\cosh\left(\lambda^{2}t\right)\right]\left[I_{2}\left(t\right)-I_{3}\left(t\right)\right]\mathrm{Re}\left\{ r_{12}\left(t\right)\right\} ,\\
\frac{d}{dt}\mathrm{Im}\left\{ r_{12}\left(t\right)\right\} = & -4\eta\left[\sinh\left(\lambda^{2}t\right)+\cosh\left(\lambda^{2}t\right)\right]\left[I_{2}\left(t\right)+I_{3}\left(t\right)\right]\mathrm{Im}\left\{ r_{12}\left(t\right)\right\} .
\end{cases}
\]

Simplifying Eqs. (\ref{I2aprox}) and (\ref{I3aprox}) gives

\[
\begin{cases}
\frac{d}{dt}\mathrm{Re}\left\{ r_{12}\left(t\right)\right\} = & -4\eta\lambda^{2}\left[2g_{0}-e^{-2\lambda^{2}t}g_{1}\left(t\right)+e^{-2\lambda^{2}t}g_{2}\left(t\right)\right]\mathrm{Re}\left\{ r_{12}\left(t\right)\right\} ,\\
\frac{d}{dt}\mathrm{Im}\left\{ r_{12}\left(t\right)\right\} = & 4\eta\lambda^{2}\left[2g_{0}-e^{2\lambda^{2}t}g_{1}\left(t\right)-e^{2\lambda^{2}t}g_{2}\left(t\right)\right]\mathrm{Im}\left\{ r_{12}\left(t\right)\right\} ,
\end{cases}
\]
whose solutions, in terms of the initial instant $t_{0}=0$, are given
by

\[
\begin{cases}
\mathrm{Re}\left\{ r_{12}\left(t\right)\right\} = & \mathrm{Re}\left\{ \rho_{12}^{\left(z\right)}\left(0\right)\right\} e^{-8\eta\lambda^{2}g_{0}t}\exp\left\{ 4\eta\lambda^{2}\int_{0}^{t}e^{-2\lambda^{2}t'}g_{1}\left(t'\right)dt'\right\} \\
 & \times\exp\left\{ -4\eta\lambda^{2}\int_{0}^{t}e^{-2\lambda^{2}t'}g_{2}\left(t'\right)dt'\right\} ,\\
\\
\mathrm{Im}\left\{ r_{12}\left(t\right)\right\} = & \mathrm{Im}\left\{ \rho_{12}^{\left(z\right)}\left(0\right)\right\} e^{8\eta\lambda^{2}g_{0}t}\exp\left\{ -4\eta\lambda^{2}\int_{0}^{t}e^{2\lambda^{2}t'}g_{1}\left(t'\right)dt'\right\} \\
 & \times\exp\left\{ -4\eta\lambda^{2}\int_{0}^{t}e^{2\lambda^{2}t'}g_{2}\left(t'\right)dt'\right\} .
\end{cases}
\]
Calculating the integrals \cite{key-43} results in Eq. (\ref{sxcoeraprox}).

\section*{Appendix G: Calculation of the trace in Eq. (\ref{eq:algorithm2})}

A convenient method for calculating the partial trace in Eq. (\ref{eq:algorithm2})
involves coherent states. For any sequence of $q_{1},\ldots,q_{N}$,
this partial trace takes the form of a matrix element of the vacuum:

\begin{eqnarray*}
\mathrm{Tr}_{B}\left\{ \prod_{n=1}^{N}\left[\hat{\hat{K}}_{q_{n}}\left(\Delta t\right)\right]\left|0\right\rangle \left\langle 0\right|\right\}  & = & \prod_{k}\left\langle 0\right|_{k}\prod_{n=0}^{N-1}\left[e^{i\omega_{k}\left(\hat{b}_{k}-q_{N-n}g_{k}/\omega_{k}\right)^{\dagger}\left(\hat{b}_{k}-q_{N-n}g_{k}/\omega_{k}\right)\Delta t}\right]\\
 & \times & \prod_{n=1}^{N}\left[e^{-i\omega_{k}\left(\hat{b}_{k}+q_{n}g_{k}/\omega_{k}\right)^{\dagger}\left(\hat{b}_{k}+q_{n}g_{k}/\omega_{k}\right)\Delta t}\right]\left|0\right\rangle _{k}.
\end{eqnarray*}
The exponential operators present in this matrix element are best
represented by displacement operators $\hat{D}_{k}\left(g_{k}/\omega_{k}\right)$
from quantum optics, which allow us to rewrite the partial trace as:

\begin{eqnarray*}
\mathrm{Tr}_{B}\left\{ \prod_{n=1}^{N}\left[\hat{\hat{K}}_{q_{n}}\left(\Delta t\right)\right]\left|0\right\rangle \left\langle 0\right|\right\}  & = & \prod_{k}\left\langle 0\right|_{k}\prod_{n=0}^{N-1}\left[\hat{D}\left(q_{N-n}\frac{g_{k}}{\omega_{k}}\right)e^{i\omega_{k}\hat{b}_{k}^{\dagger}\hat{b}_{k}\Delta t}\hat{D}^{\dagger}\left(q_{N-n}\frac{g_{k}}{\omega_{k}}\right)\right]\\
 & \times & \prod_{n=1}^{N}\left[\hat{D}^{\dagger}\left(q_{n}\frac{g_{k}}{\omega_{k}}\right)e^{-i\omega_{k}\hat{b}_{k}^{\dagger}\hat{b}_{k}\Delta t}\hat{D}\left(q_{n}\frac{g_{k}}{\omega_{k}}\right)\right]\left|0\right\rangle _{k}.
\end{eqnarray*}
Any sequence of these operators applied to a coherent state $\left|\alpha_{k}\right\rangle $
yields:

\[
\left[\hat{D}^{\dagger}\left(q_{n}\frac{g_{k}}{\omega_{k}}\right)e^{-i\omega_{k}\hat{b}_{k}^{\dagger}\hat{b}_{k}\Delta t}\hat{D}\left(q_{n}\frac{g_{k}}{\omega_{k}}\right)\right]\left|\alpha_{k}\right\rangle =\left|e^{-i\omega_{k}\Delta t}\alpha_{k}+\left(e^{-i\omega_{k}\Delta t}-1\right)q_{n}\frac{g_{k}}{\omega_{k}}\right\rangle ,
\]
where we have discarded the complex phase factors due to the displacement
operators. Repeating the procedure $N$ times, we find that the partial
trace is simply the inner product of coherent states:

\[
\prod_{k}\left\langle \sum_{n=1}^{N}e^{-in\omega_{k}\Delta t}\left(e^{i\omega_{k}\Delta t}-1\right)q_{N-n+1}\frac{g_{k}}{\omega_{k}}\right.\left|\sum_{n=1}^{N}e^{-in\omega_{k}\Delta t}\left(1-e^{i\omega_{k}\Delta t}\right)q_{N-n+1}\frac{g_{k}}{\omega_{k}}\right\rangle .
\]
Such an inner product results in the exponential:

\begin{eqnarray*}
\mathrm{Tr}_{B}\left\{ \prod_{n=1}^{N}\left[\hat{\hat{K}}_{q_{n}}\left(\Delta t\right)\right]\left|0\right\rangle \left\langle 0\right|\right\}  & = & \exp\left\{ -8\sum_{m=1}^{N}\sum_{n=1}^{N}q_{m}q_{n}\int_{0}^{\infty}d\omega\;\frac{J\left(\omega\right)}{\omega^{2}}\sin^{2}\left(\frac{\omega\Delta t}{2}\right)\right.\\
 & \times & \left.\cos\left[\left(m-n\right)\omega\Delta t\right]\right\} ,
\end{eqnarray*}
where we have taken the limit to a continuous spectrum of frequencies,
applying the definition of the spectral-density function given in
Eq. (\ref{eq:Spectraldensity}).

For Ohmic spectral densities, Eq. (\ref{eq:OhmicSD}), the partial
trace becomes the following exponential of an integral:

\begin{eqnarray*}
\mathrm{Tr}_{B}\left\{ \prod_{n=1}^{N}\left[\hat{\hat{K}}_{q_{n}}\left(\Delta t\right)\right]\left|0\right\rangle \left\langle 0\right|\right\}  & = & \prod_{m=1}^{N}\prod_{n=1}^{N}\exp\left\{ -8\eta q_{m}q_{n}\int_{0}^{\infty}d\omega\frac{e^{-\omega/\omega_{c}}}{\omega}\sin^{2}\left(\frac{\omega\Delta t}{2}\right)\right.\\
 & \times & \left.\cos\left[\left(m-n\right)\omega\Delta t\right]\right\} .
\end{eqnarray*}
The remaining integral in $d\omega$ may be solved by first noticing
that $\int_{0}^{\Delta t}d\tau\;\sin\left(\omega\tau\right)=\frac{2}{\omega}\sin^{2}\left(\frac{\omega\Delta t}{2}\right)$.
This leads to a double integral that is easily solved as:

\begin{eqnarray*}
4\int_{0}^{\Delta t}d\tau\int_{0}^{\infty}d\omega\; e^{-\omega/\omega_{c}}\sin\left(\omega\tau\right)\cos\left[\left(m-n\right)\omega\Delta t\right] & = & \ln\left\{ 1+\frac{2\left(\omega_{c}\Delta t\right)^{2}}{\left[1+\left(m-n\right)^{2}\left(\omega_{c}\Delta t\right)^{2}\right]^{2}}\right.\\
 & + & \left.\frac{\left[1-2\left(m-n\right)^{2}\right]\left(\omega_{c}\Delta t\right)^{4}}{\left[1+\left(m-n\right)^{2}\left(\omega_{c}\Delta t\right)^{2}\right]^{2}}\right\} ,
\end{eqnarray*}
which gives the following result for the trace:

\[
\mathrm{Tr}_{B}\left\{ \prod_{n=1}^{N}\left[\hat{\hat{K}}_{q_{n}}\left(\Delta t\right)\right]\left|0\right\rangle \left\langle 0\right|\right\} =\prod_{m=1}^{N}\prod_{n=1}^{N}\left\{ 1+\frac{2\left(\omega_{c}\Delta t\right)^{-2}+\left[1-2\left(m-n\right)^{2}\right]}{\left[\left(\omega_{c}\Delta t\right)^{-2}+\left(m-n\right)^{2}\right]^{2}}\right\} ^{-\eta q_{m}q_{n}}.
\]


\begin{thebibliography}{References}
\bibitem{key-1}C. Cohen-Tannoudji, B. Diu and F. Laloë, \emph{Quantum
mechanics} (Wiley, New York, 1977).

\bibitem{key-2}J. A. Wheeler and W. H. Zurek, \emph{Quantum theory
and measurement} (Princeton University Press, Princeton, 1983).

\bibitem{key-3}J. von Neumann, \emph{Mathematical foundations of
quantum mechanics} (Princeton University Press, Princeton, 1955).

\bibitem{key-4}P. A. M. Dirac, \emph{The principles of quantum mechanics}
(Clarendon Press, Oxford, 1958).

\bibitem{key-31}A. Bassi and G. Ghirardi, Phys. Rep. 379, 257 (2003).

\bibitem{key-44}B. L. van der Waerden, \emph{Sources of quantum mechanics}
(North-Holland, Amsterdam, 1967).

\bibitem{key-45}G. Ludwig, \emph{Wave mechanics} (Pergamon Press,
Oxford, 1968).

\bibitem{key-46}P. A. M. Dirac, Proc. R. Soc. Lond. A 112, 661 (1926).

\bibitem{key-5}A. Peres, Phys. Rev. A 61, 022116 (2000).

\bibitem{key-6}S. L. Adler, Phys. Lett. A 265, 58 (2000).

\bibitem{key-7}G. Lindblad, Commun. Math. Phys. 48, 119 (1976).

\bibitem{key-33}E. B. Davies, \emph{Quantum theory of open systems}
(Academic Press, London, 1976).

\bibitem{key-34}R. Alicki and K. Lendi, \emph{Quantum dynamical semigroups
and applications} (Springer-Verlag, Berlin, 2007).

\bibitem{key-8}I. Percival, \emph{Quantum state diffusion} (Cambridge
University Press, Cambridge, 1998).

\bibitem{key-9}H. P. Breuer and F. Petruccione, \emph{The theory
of open quantum systems} (Oxford University Press, Oxford, 2002).

\bibitem{key-11}S. Nakajima, Progr. Theor. Phys. 20, 948 (1958).

\bibitem{key-12}R. Zwanzig, J. Chem. Phys. 33, 1338 (1960).

\bibitem{key-10}C. A. Brasil and R. d. J. Napolitano, e-print arXiv:1102.3667v2.

\bibitem{key-16}J. Fischer and H. P. Breuer, Phys. Rev. A 76, 052119
(2007).

\bibitem{key-36}H. P. Breuer, Phys. Rev. A, 75, 022103 (2007).

\bibitem{key-17}J. Seke, J. Phys. A: Math. Gen., 23, L61 (1990).

\bibitem{key-18}A. Smirne and B. Vacchini, Phys. Rev. A 82, 022110
(2010).

\bibitem{key-19}W. T. Strunz, L. Diosi and N. Gisin, Phys. Rev. Lett.
82, 1801 (1999).

\bibitem{key-39}K. Dietz, J. Phys. A: Math. Gen., 36, L45 (2003).

\bibitem{key-41}W. H. Press, S. A. Teukolsky, W. T. Vetterling and
B. P. Flannery, \emph{Numerical Recipes - The Art of Scientific Computing}
(Cambridge University Press, New York, 2007).

\bibitem{key-20}C. R. Willis and R. H. Picard, Phys. Rev. A 9, 1343
(1974).

\bibitem{key-30}R. H. Picard and C. R. Willis, Phys. Rev. A 16, 1625
(1977).

\bibitem{key-37}S. Jang, J. Cao and R. J. Silbey, J. Chem. Phys.
116, 2705 (2002).

\bibitem{key-38}S. Jang, J. Cao and R. J. Silbey, J. Chem. Phys.
117, 10428 (2002).

\bibitem{key-40}A. Peres and W. K. Wootters, Phys. Rev. D 32, 1968
(1985).

\bibitem{key-42}R. Wilcox, J. Math. Phys. 8, 962 (1967).

\bibitem{key-43}I.S. Gradshteyn and I.M. Ryzhik, \emph{Table of Integrals,
Series, and Products 7th Edition} (Elsevier-Academic Press, Burlington,
2007).

\bibitem{key-13}A. Peres, Am. J. Phys. 48, 931 (1980).

\bibitem{key-14}B. Misra and E. C. G. Sudarshan, J. Math. Phys. 18,
756 (1977).

\bibitem{key-15}C. B. Chiu, E. C. G. Sudarshan and B. Misra, Phys.
Rev. D 16, 520 (1977).

\bibitem{key-47}P. Facchi and S. Pascazio, Phys. Rev. Lett. 89, 080401
(2002).

\bibitem{key-48}P. Facchi, D. A. Lidar and S. Pascazio, Phys. Rev.
A 69, 032314 (2004).\end{thebibliography}
\end{document}